\documentclass[%
reprint,
superscriptaddress,
bibnotes,
amsmath,amssymb,
aps,
prl,
nolongbibliography,
]{revtex4-2}

\usepackage{graphicx}
\usepackage{dcolumn}
\usepackage{bm}
\usepackage{chemformula}
\usepackage{hyperref}

\usepackage{tabularx}
\usepackage{amsmath}
\usepackage{float}
\usepackage{booktabs}
\usepackage{animate}
\usepackage{multirow}

\begin{document}


\title{Magnetic molecular orbitals in {MnSi}}

\author{Zhendong~Jin}
\thanks{These authors contributed equally to this study.}
\affiliation{International Center for Quantum Materials, School of Physics, Peking University, Beijing 100871, China}
\author{Yangmu~Li}
\thanks{These authors contributed equally to this study.}
\affiliation{Beijing National Laboratory for Condensed Matter Physics, and Institute of Physics, Chinese Academy of Sciences, Beijing 100190, China}
\affiliation{Condensed Matter Physics and Materials Science Division, Brookhaven National Laboratory, Upton, New York 11973, USA}
\affiliation{University of Chinese Academy of Sciences, Beijing 100049, China}
\author{Zhigang~Hu}
\thanks{These authors contributed equally to this study.}
\affiliation{International Center for Quantum Materials, School of Physics, Peking University, Beijing 100871, China}
\author{Biaoyan~Hu}
\affiliation{International Center for Quantum Materials, School of Physics, Peking University, Beijing 100871, China}
\author{Yiran~Liu}
\affiliation{International Center for Quantum Materials, School of Physics, Peking University, Beijing 100871, China}
\author{Kazuki~Iida}
\affiliation{Neutron Science and Technology Center, Comprehensive Research Organisation for Science and Society (CROSS), Tokai, Ibaraki 319-1106, Japan}
\author{Kazuya~Kamazawa}
\affiliation{Neutron Science and Technology Center, Comprehensive Research Organisation for Science and Society (CROSS), Tokai, Ibaraki 319-1106, Japan}
\author{M.~B.~Stone}
\affiliation{Neutron Scattering Division, Oak Ridge National Laboratory, Oak Ridge, Tennessee 37831, USA}
\author{A.~I.~Kolesnikov}
\affiliation{Neutron Scattering Division, Oak Ridge National Laboratory, Oak Ridge, Tennessee 37831, USA}
\author{D.~L.~Abernathy}
\affiliation{Neutron Scattering Division, Oak Ridge National Laboratory, Oak Ridge, Tennessee 37831, USA}
\author{Xiangyu~Zhang}
\affiliation{State Key Laboratory for Advance Metals and Materials, University of Science and Technology Beijing, Beijing 10083, China}
\author{Haiyang~Chen}
\affiliation{State Key Laboratory for Advance Metals and Materials, University of Science and Technology Beijing, Beijing 10083, China}
\author{Yandong~Wang}
\affiliation{State Key Laboratory for Advance Metals and Materials, University of Science and Technology Beijing, Beijing 10083, China}
\author{Chen Fang}
\affiliation{Beijing National Laboratory for Condensed Matter Physics, and Institute of Physics, Chinese Academy of Sciences, Beijing 100190, China}
\author{Biao~Wu}
\affiliation{International Center for Quantum Materials, School of Physics, Peking University, Beijing 100871, China}
\author{I.~A.~Zaliznyak}
\affiliation{Condensed Matter Physics and Materials Science Division, Brookhaven National Laboratory, Upton, New York 11973, USA}
\author{J.~M.~Tranquada}
\affiliation{Condensed Matter Physics and Materials Science Division, Brookhaven National Laboratory, Upton, New York 11973, USA}
\author{Yuan~Li}
\email{yuan.li@pku.edu.cn}
\affiliation{International Center for Quantum Materials, School of Physics, Peking University, Beijing 100871, China}
\affiliation{Tsung-Dao Lee Institute and School of Physics \& Astronomy, Shanghai Jiao Tong University, Shanghai 200240, China}

\date{\today}

\begin{abstract}
A large body of knowledge about magnetism is attained from models of interacting spins, which usually reside on magnetic ions. Proposals beyond the ionic picture are uncommon and seldom verified by direct observations in conjunction with microscopic theory. Here, using inelastic neutron scattering to study the itinerant near-ferromagnet MnSi, we find that the system's fundamental magnetic units are interconnected, extended molecular orbitals consisting of three Mn atoms each, rather than individual Mn atoms. This result is further corroborated by magnetic Wannier orbitals obtained by \textit{ab initio} calculations. It contrasts the ionic picture with a concrete example, and presents a novel regime of the spin waves where the wavelength is comparable to the spatial extent of the molecular orbitals. Our discovery brings important insights into not only the magnetism of MnSi, but also a broad range of magnetic quantum materials where structural symmetry, electron itinerancy and correlations act in concert.

\end{abstract}

\maketitle

It is intuitive to use networks of interacting spins to describe magnetism in insulators, yet even for itinerant systems such as semiconductors \cite{SatoRMP2010} and metals \cite{KvashninPRL2016}, the notion of interacting moments sitting on real-space lattice sites remains useful.
Spins are usually identified with magnetic ions, which corresponds to tight-binding (TB) electronic models constructed from atomic orbitals (AOs).
But the ionic picture is not always accurate. Pauling first pointed out \cite{PaulingJACS1931} that covalent bonding alters transition metals' electron count and moment compared to the ionic case. A modern insight into the problem can be obtained by considering the notion of Wannier orbitals, computable as maximally localized Wannier functions (MLWFs) \cite{MarzariRMP2012}, which replace AOs and form a natural basis for low-energy TB models \cite{PoPRX2018,KoshinoPRX2018,KangPRX2018}.
They also describe magnetic electron clouds with unpaired spins \cite{MarzariRMP2012,
WaltersNatPhys2009}. While exponentially localized and symmetric Wannier orbitals generally exist in ``topologically trivial'' materials \cite{BradlynNature2017,PoNatComm2017
}, they are not always centered on atomic sites \cite{Xu2021,Gao2021,Li2021}. For instance, when tightly-bonded clusters of atoms are embedded in a loosely-bonded matrix \cite{Sheckelton2012,Kim2014,
Khomskii2021
}, electrons can become partially delocalized on the clusters but have weak inter-cluster hopping, forming molecular orbitals (MOs). Magnetic MOs are recently considered a fertile ground for finding new quantum magnets \cite{Nguyen2021}.

The formation of MOs can also be an emergent phenomenon without a cluster-like structural motif. In such cases, MOs can be mathematically distinguished by their off-atom Wannier centers \cite{Xu2021,Gao2021}, of which an instructive (non-magnetic) example is silicon, where the valence electrons are centered on Si-Si bonds \cite{MarzariPRB1997,Gao2021,Li2021}. Such interconnected MOs are characteristic of electron delocalization in covalent solids, to which many transition-metal compounds actually belong, at least partially \cite{WaltersNatPhys2009,TomiyasuPRL2014}. Indeed, when a correlated metal's electronic interaction strength is continuously tuned towards a Mott transition, one can imagine a state where MOs form out of a balance between electronic interactions and kinetic energy. To give some examples, benzene-like MOs have been theoretically proposed in $4d$ and $5d$ transition-metal oxides with a honeycomb lattice \cite{MazinPRL2012,FoyevtsovaPRB2013,StreltsovPRB2015}, challenging some of the materials' potential to realize the Kitaev model \cite{Takagi19}. Orbital loop currents proposed in cuprate \cite{VarmaPRB1997} and kagome \cite{FengSB2021} superconductors can be viewed as MOs with unquenched orbital magnetic moments. Recent neutron scattering experiments in iron chalcogenide superconductors have indicated presence of unusual magnetic plaquettes \cite{ZaliznyakPNAS2015}, which are consistent with magnetic MOs despite their formation mechanism is not clearly understood \cite{TranquadaJPCM2020}. Establishing an emergent magnetic MO reality can thus be expected to fundamentally influence research in a broad range of correlated-electron systems.

Here we present a discovery of interconnected emergent magnetic MOs in a simple binary compound MnSi using inelastic neutron scattering (INS). The central idea stems from the concept of magnetic form factor, which is a momentum-dependent prefactor of the magnetic neutron scattering cross section \cite{Shirane2002}, given by the Fourier transform of the electron cloud associated with each spin. The key is that a magnetic MO's form factor has distinct structures that can be experimentally observed. MnSi hosts intriguing physics including chiral magnetism \cite{MuhlbauerScience2009} and electronic nodal planes \cite{JeongPRB2004,WildeNature2021} due to its non-centrosymmetric and non-symmorphic crystal structure (fig.~\ref{Figs1} \cite{SM}), yet its advantage for our purpose is its itinerant magnetism \cite{IshikawaPRB1977,MuhlbauerScience2009}: Magnetic excitations of MnSi are well-characterized \cite{IshikawaPRB1977}, hence allowing for a definitive measurement of the magnetic form factor \cite{WaltersNatPhys2009,TomiyasuPRL2014}. The electronic structure can be reliably calculated with density functional theory (DFT) \cite{JeongPRB2004,NicolaouPRB2015
}, which enables calculation of the Wannier orbitals. Combined together, our results conclusively show that the fundamental magnetic units of MnSi are extended MOs and not AOs of individual Mn.
\begin{figure}
	\centering{\includegraphics[clip,width=8.5cm]{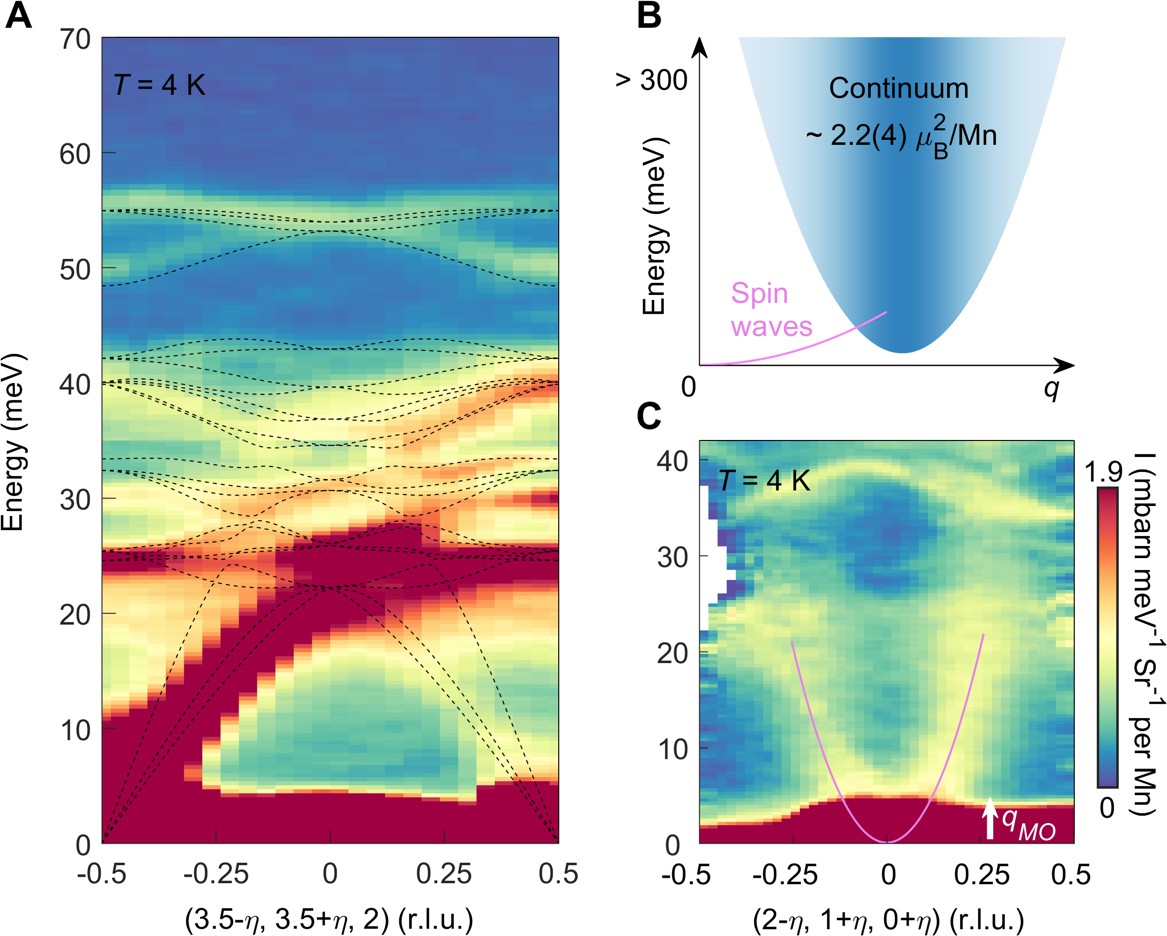}}
	\caption{\textbf{Overview of INS signals.} (\textbf{A}) Phonons measured in high-index BZs. DFT-calculated dispersions are shown as dashed lines. (\textbf{B}) Schematic of magnetic excitations, including spin waves and a continuum up to high energies with substantial spectral weight (estimated without form-factor reduction \cite{SM}). (\textbf{C}) Spin waves in the $(2, 1, 0)$ BZ. The signal becomes weak beyond $q_{\mathrm{MO}} \approx 0.65\,\mathrm{\AA}^{-1}$ from the zone center (figs.~\ref{Figs4} and \ref{Figs5} \cite{SM}). Phonons are visible above $\sim 20$ meV (fig.~\ref{Figs4} \cite{SM}). Magenta lines are spin-wave dispersion \cite{IshikawaPRB1977}.}
\label{Fig1}
\end{figure}

To begin, we present in Fig.~\ref{Fig1}A our INS measurement \cite{SM} of phonons using a 33-gram twin-free single-crystal sample (fig.~\ref{Figs2} \cite{SM}). DFT-calculated phonon dispersions \cite{SM} are overlaid with the data and show very good agreement. With this information, we can then identify magnetic signals by avoiding the phonons in momentum ($\mathbf{Q}$) and energy ($E$) transfer. Figure~\ref{Fig1}B illustrates the two types of magnetic signals we observe. First, at high energy there is a continuum of excitations carrying a significant spectral weight. The wide energy span ($\geq$ 300 meV)
of the excitations suggests that they are related to conduction electrons, whose band widths are over 1~eV \cite{JeongPRB2004,NicolaouPRB2015}. Second, at low energies and emanating from the Brillouin zone (BZ) center, there are (nearly) ferromagnetic spin waves with a parabolic dispersion \cite{JanoschekPRB2010}. These modes are damped because of coupling to the conduction electrons \cite{IshikawaPRB1977}. Consistent with a recent report \cite{Chen2020}, we observe the spin waves in the BZ of $(2, 1, 0)$ (in reciprocal lattice units, r.l.u.), where phonon signals are weak (Fig.~\ref{Fig1}C). The spin-wave dispersion is distinct from the phonons and can be observed in other BZs (fig.~\ref{Figs3} \cite{SM}). As has been previously found \cite{IshikawaPRB1977}, both the spin waves (paramagnons) and the continuum persist far above the magnetic ordering temperature of 29.5~K (fig.~\ref{Figs7} and \ref{Figs8} \cite{SM}).

\begin{figure}
	\centering{\includegraphics[clip,width=8.5cm]{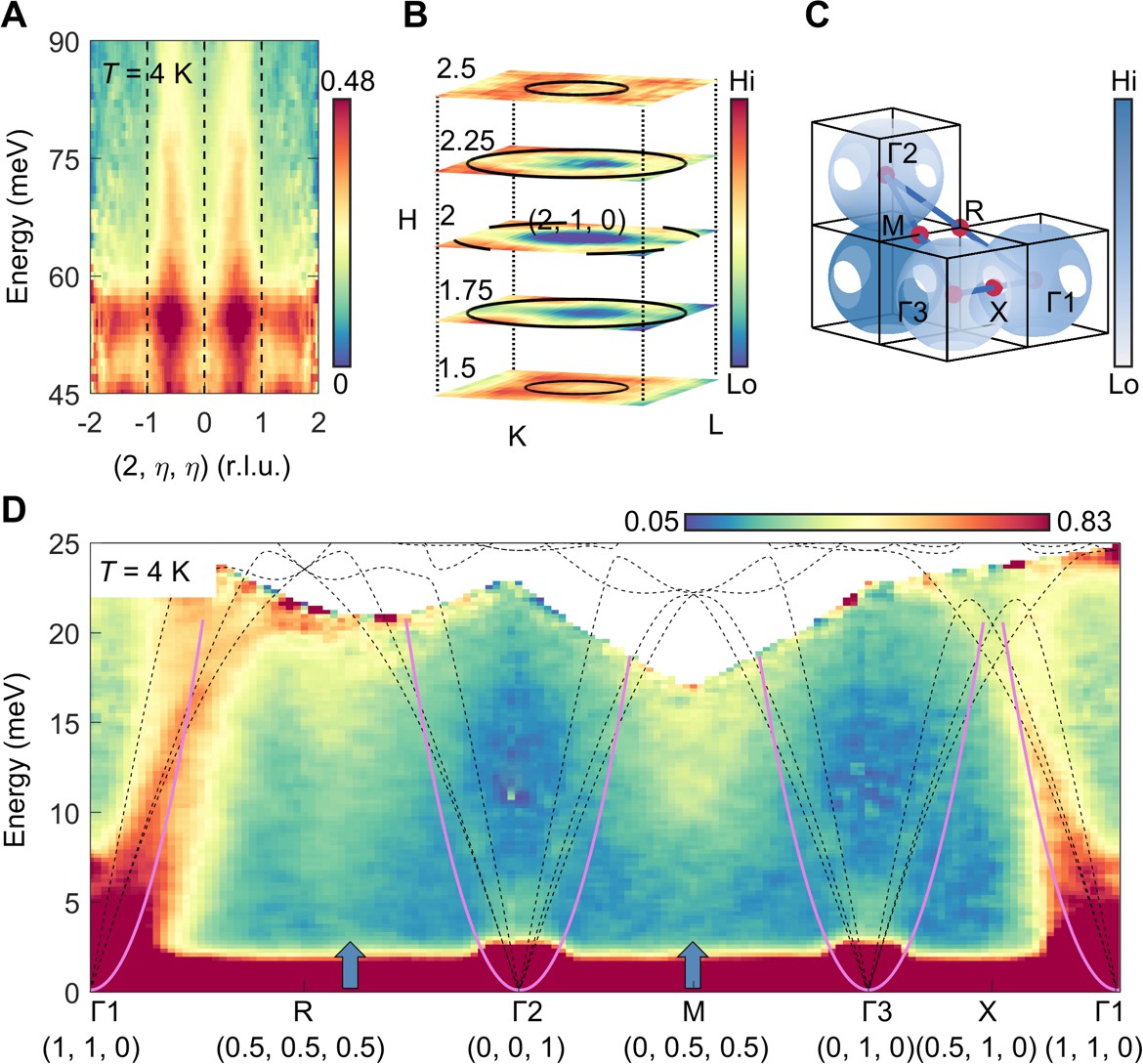}}
	\caption{\textbf{Momentum structure of excitation continuum.} (\textbf{A}) Nearly energy-independent spectra above phonon energies. (\textbf{B}) Intra-BZ structure measured at $70\pm10$ meV in the $(2, 1, 0)$ BZ, normalized in each constant-$H$ slice. Black circles represent a sphere of radius 0.76 ${\rm\AA^{-1}}$ (``shell model'', see text). (\textbf{C}) Connected spherical shells over several BZs. Calculated scattering cross section (Fig.~\ref{Fig3}) is displayed in gray scale. (\textbf{D}) Low-energy spectra along the $\mathbf{Q}$ trajectory indicated in (C). The bottom of the continuum (arrows) is seen near the M- and R-points, below the phonon (black dotted lines) and spin-wave (magenta lines) dispersions.}
\label{Fig2}
\end{figure}
\begin{figure*}
	\centering{\includegraphics[clip,width=17cm]{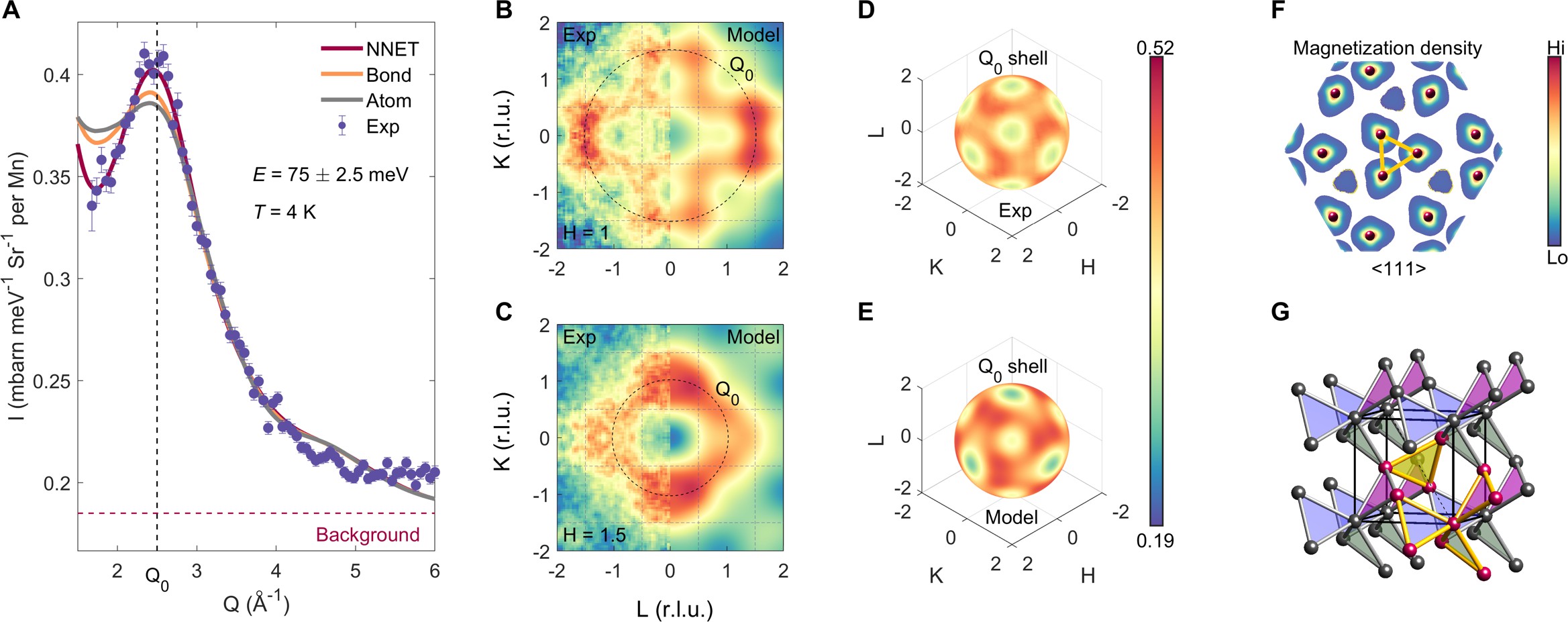}}
\caption{\textbf{Model for magnetic form factor.} (\textbf{A}) Radial intensity distribution of the continuum, after averaging over all directions in $\mathbf{Q}$. $Q_{\rm 0} = 2.5\,{\rm\AA^{-1}}$ indicates a local maximum. Magnetic clusters centered on NNETs, Mn-Mn bonds, and Mn atoms are used in conjunction with the shell model to simulate the intensity (see text and figs.~\ref{Figs12} to \ref{Figs14} \cite{SM} without the shell model), and their optimal results are compared with the data. Error bars are statistical uncertainty 1 s.d. (\textbf{B}-\textbf{E}) INS data in comparison with model calculations (see text and \cite{SM}). Experimental results are obtained at $E = 75 \pm 10$ meV, $T$ = 4 K. (\textbf{F}) DFT-calculated magnetization densities in an atomic plane perpendicular to $\langle111\rangle$. Three neighboring Mn atoms are connected by a NNET (fig.~\ref{Figs1} \cite{SM}). (\textbf{G}) The Mn sublattice, organized into NNETs with their four orientations indicated by colours. Highlighted atoms and bonds indicate a ``trillium flower'' \cite{HopkinsonPRB2006} structural motif.
}
	\label{Fig3}
\end{figure*}

Figure~\ref{Fig2}A displays the excitation continuum at energies higher than the phonon bands. The observed $\mathbf{Q}$-space structures are nearly independent of $E$ (up to at least 250 meV, see figs.~\ref{Figs6} and \ref{Figs7} \cite{SM}), presumably because the measured energy window is narrow compared to the conduction band width. The continuum's intensity clearly varies between the BZs, and within the BZ of $(2, 1, 0)$, it can be approximately described by a thick spherical shell (``shell model'' hereafter) of a diameter slightly greater than 1 r.l.u. (Fig.~\ref{Fig2}B). Shells belonging to adjacent BZs are merged together, leaving holes around the X-points of the BZ (Fig.~\ref{Fig2}C). The shell model is also consistent with signals coming from the energy minimum of the continuum, which we observe as broad features near the R- and M-points in Fig.~\ref{Fig2}D, below the phonon and the spin-wave dispersions. In fig.~\ref{Figs9} \cite{SM}, we show that this low-energy signal varies across the BZs in the same way as the high-energy continuum. In contrast, the spin waves and the continuum have different intensity variations across the BZs (Fig.~\ref{Fig1}C and Fig.~\ref{Fig2}D, see also fig.~\ref{Figs10} \cite{SM}). Together with the fact that the continuum starts from an energy below the top of the well-defined spin-wave dispersion, these results indicate that the two are distinct types of excitations, rather than belonging to the same physical entity as recently suggested \cite{Chen2020}.

We now use the diffuse and dispersion-less continuum as a probe of the magnetic form factor, which we expect to be encoded in the inter-BZ intensity variations. The wide energy bandwidth enables us to access a wide $\mathbf{Q}$ range and corroborate the data consistency (figs.~\ref{Figs6}, \ref{Figs7} and \ref{Figs11} \cite{SM}). Figure~\ref{Fig3}A presents the radial $Q$ dependence of the signal after averaging over all directions in $\mathbf{Q}$. Surprisingly, the intensity exhibits a distinct local maximum at $Q_0 = 2.5\,\mathrm{\AA}^{-1}$, rather than being a monotonically decreasing function of $Q$ as expected for the atomic form factor \cite{Shirane2002}. This local maximum immediately suggests that the magnetization cloud has an internal characteristic length of $\sim2 \pi/Q_0$. The intensity exhibits rich inter-BZ variations (Fig.~\ref{Fig3}, B and C), including a pronounced directional dependence (Fig.~\ref{Fig3}D) on the momentum sphere $|\mathbf{Q}| = Q_0$.

Despite the apparent complexity, \textit{all} of the above $\mathbf{Q}$ dependence can be quantitatively captured by a simple and intuitive model for the magnetic form factor. To introduce the model, we note that the DFT-calculated magnetization density (Fig.~\ref{Fig3}F) is concentrated near the Mn atoms, which can be organized into nearest-neighbor equilateral triangles (NNETs) perpendicular to the four $\langle111\rangle$ directions (Fig.~\ref{Fig3}G and fig.~\ref{Figs1} \cite{SM}). This suggests a model where individual NNETs are the fundamental magnetic units. In our model, each spin is assumed to reside equally on all three vertices of a NNET (side $=2.8$~\AA, close to $2 \pi/Q_0$). By computing the form factor of the NNET \cite{SM} averaged over the four orientations (Fig.~\ref{Fig3}G) and using the result in the shell model (Fig.~\ref{Fig2}C), we obtain an excellent description of the $\mathbf{Q}$-dependent intensities of the continuum (Fig.~\ref{Fig3}, A to E). The only fit parameters are a background constant, an overall intensity scale, and the diameter and Gaussian thickness of the intra-BZ shells. The inclusion of the shell model has nearly no effect on the calculation in Fig.~\ref{Fig3}A (NNET), and it only adds periodic fine details to the results in Fig.~\ref{Fig3}, B, C and E. The inter-BZ variations entirely arise from the NNET form factor (fig.~\ref{Figs15} \cite{SM}), which has no adjustable parameter and very little dependence on the description of the magnetization densities on Mn (fig.~\ref{Figs16} \cite{SM}). To assess whether the form factor associated with the NNET cluster is uniquely favored, alternative clusters of different sizes have been tested (figs.~\ref{Figs12} to \ref{Figs14} \cite{SM}), and none of them can describe the data as well as the NNET (Fig.~\ref{Fig3}A). Indeed, the NNET form factor is able to describe the \textit{global} intensity variations of the continuum, down to its energy minimum and into the first BZ (Fig.~\ref{Fig2}, C, D and figs.~\ref{Figs16}, \ref{Figs17} \cite{SM}).

\begin{figure}
	\centering{\includegraphics[clip,width=8.5cm]{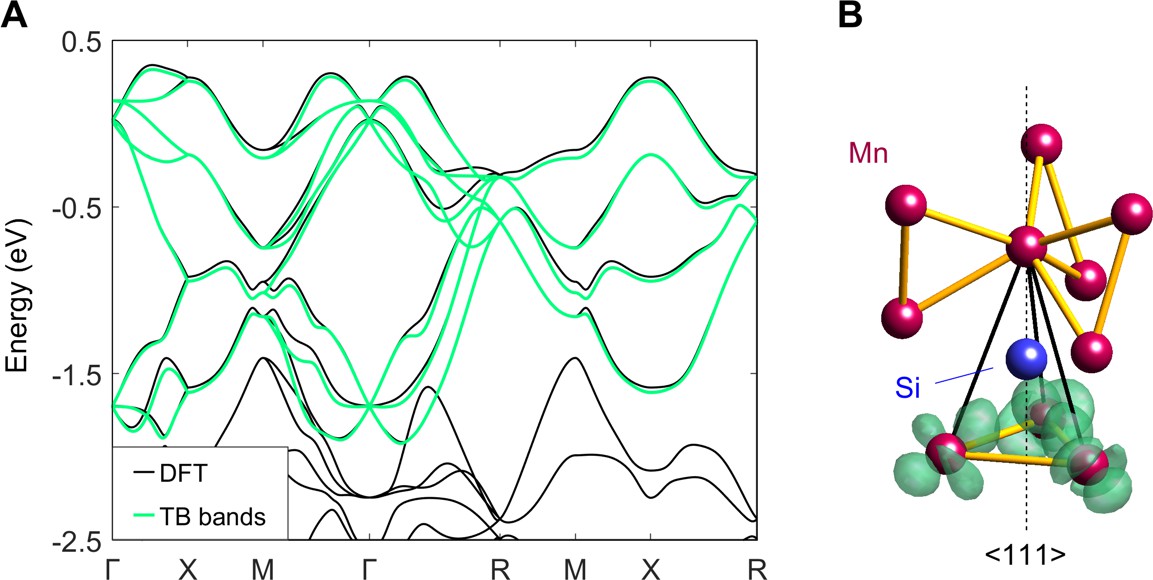}}
	\caption{
\textbf{\textit{Ab initio} calculation of magnetic orbitals.} (\textbf{A}) DFT-calculated bands near $E_\mathrm{F}$ in comparison to TB models. For clarity, only spin-majority bands are shown (minority bands in fig.~\ref{Figs23} \cite{SM}). (\textbf{B}) MLWFs calculated from the EBR, resulting in MOs on NNETs.  Isosurface (15\% of maximum) of the electron density is shown on a structural motif highlighted in Fig.~\ref{Fig3}G. The spread of electron density $\left\langle r^2 \right\rangle - \left\langle r \right\rangle^2$ \cite{MarzariRMP2012} is 11.3~$\mathrm{\AA}^2$. The Mn atom above Si has AFM coupling \cite{ShanavasPRB2016} to those at the bottom, which translates into AFM coupling between neighboring NNETs (Fig.~\ref{Fig3}G).}
	\label{Fig4}
\end{figure}

The success of the NNET form factor indicates that the individual magnetization clouds are off-atom MOs. These MOs result from the hybridization of magnetic electrons' wave functions and are distinct from magnetic correlations between neighboring Mn atoms enforced by ferromagnetic interactions -- the strength of the interactions would have to be unrealistically large to support dynamic correlations at hundreds of meV. To elucidate the electronic structure behind the MOs, we examine the eight magnetic conduction bands (Fig.~\ref{Fig4}A) crossing the Fermi level ($E_\mathrm{F}$). Symmetry analysis (table~\ref{band representation} \cite{SM}) shows that these bands form an elementary band representation (EBR), which opens up the possibility for them to be described by a stand-alone TB model \cite{BradlynNature2017}.
We have performed a MLWF calculation restricted to the EBR. Consistent with the experiment, the obtained MLWFs are indeed found to be centered on NNETs, and the resultant TB model provides an excellent description of the magnetic bands (Fig.~\ref{Fig4}A). We note that the high electron densities are offset from the NNET Wannier center and are near the Mn atoms (Fig.~\ref{Fig4}B), which closely resembles flat-band electrons in twisted graphene \cite{KangPRX2018,PoPRX2018,KoshinoPRX2018}. The obtained Wannier orbitals represent MOs centered at the centroid of each NNET. In contrast, MLWFs obtained under Mn-site constraint for the magnetic EBR (fig.~\ref{Figs21}D \cite{SM}), or as AOs of Mn calculated without restriction to the EBR (fig.~\ref{Figs21}E \cite{SM}), feature substantially different shape of the electronic magnetization density, whose magnetic form factor is clearly different from our experimental observation (Fig.~\ref{Fig3}A). More details on the comparison of different MLWFs can be found in \cite{SM}.


Our result brings critical insight into the magnetism of MnSi. We first note an intriguing similarity between the MOs' orientational characteristics, as seen from the INS continuum (Fig.~\ref{Fig3}, D and E), to those of spin correlations seen in MnSi's non-Fermi-liquid phase under pressure \cite{PfleidererNature2004}. The energy scale of the continuum is compatible with transitions of itinerant electrons \cite{IshikawaPRB1977,JeongPRB2004,NicolaouPRB2015,WildeNature2021}, in the context of which our shell model would be an approximation for the (intra-BZ) joint density of states. The INS intensity further depends on transition matrix elements between the hybridized electronic states -- as they all belong to the same magnetic EBR, the NNET form factor constitutes an excellent approximation for the intensity (see \cite{SM} for a mathematical proof which uses the localization property of MLWFs). Yet still, the continuum's persistence into the paramagnetic state is inconsistent with simple spin-polarized bands. Moreover, by accounting for the rapid form factor decrease towards large $Q$ (fig.~\ref{Figs16} \cite{SM}), we find a magnetic spectral weight of $\geq 2 \mu_\mathrm{B}^2/\mathrm{Mn}$
 for the continuum \cite{SM}, which is too large for purely itinerant excitations and suggests a dichotomy of itinerant and local moments.

From a local-moment perspective, our result indicates that the fundamental magnetic units are MOs rather than Mn atoms \cite{ShanavasPRB2016,ChizhikovPRB2012,DmitrienkoPRL2012}. Compared to the shell model, the continuum can be nearly equally well-described by a lattice Lorentzian model (fig.~\ref{Figs18} \cite{SM}), which accounts for short-range antiferromagnetic (AFM) correlations. The MnSi structure's Wyckoff $4a$ site features a unique ``trillium'' lattice \cite{HopkinsonPRB2006}, where the proliferation of corner-sharing triangles (Fig.~\ref{Fig3}G and Fig.~\ref{Fig4}B) is expected to result in frustrated ordering in the case of AFM interactions, consistent with a low ordering temperature and reduced ordered moment per Mn \cite{HopkinsonPRB2007,IsakovPRB2008,JansonNatComm2014}. Importantly, both the Mn and the NNET sites are Wyckoff $4a$. Even though either site can form a consistent TB description of the magnetic EBR (fig.~\ref{Figs21}A \cite{SM}), our INS data unequivocally indicate that the magnetization clouds are centered on the NNETs. While neighboring Mn atoms are unlikely to have AFM coupling \cite{ChizhikovPRB2012,DmitrienkoPRL2012,HopkinsonPRB2007}, the NNETs may have, because Mn atoms connected via a relatively straight Mn-Si-Mn bond \cite{ShanavasPRB2016} belong to neighboring NNETs (Fig.~\ref{Fig3}G and Fig.~\ref{Fig4}B). Our result motivates further study of magnetic couplings and correlations between the MOs on the NNET trillium lattice.

The magnetic MOs further introduce a novel and interesting regime of the spin waves, where the wavelength becomes short and comparable to the spatial extent of the MOs. Because a MO's magnetization cloud must act together, spin waves in this regime might become ill-defined (overdamped). For an estimate, by requiring the spin waves (as plane waves) to accommodate the circumradius of a NNET into their amplitude profile's full width at half maximum, we obtain a cutoff momentum of $q_\mathrm{MO} \approx 0.65\,\mathrm{\AA}^{-1}$, which agrees well with our data in Fig.~\ref{Fig1}C and figs.~\ref{Figs4}, \ref{Figs5} \cite{SM}: propagating spin waves can hardly be observed beyond $q_\mathrm{MO}$, where the scattering is dominated by the continuum. Whether or not the similarity between $q_\mathbf{MO}$ and the location of the continuum (Fig.~\ref{Fig2}C) is coincidental, they have very different physical origins (as already discussed, the two signals are distinct, contrary to Ref.~\cite{Chen2020}). Instead of becoming overdamped, the spin waves might also pick up an extremely large dispersion velocity beyond $q_\mathrm{MO}$ because the MOs are highly rigid, but the signal's overlap with the continuum precludes a definitive answer at this point.

To summarize, we have determined the electronic magnetization clouds in MnSi and discovered that they form magnetic MOs. The INS method works exceptionally well when the periodic intra-BZ and the global inter-BZ structures have separable momentum scales. Our concrete demonstration of magnetic MOs calls for an in-depth examination of this concept in many interesting but not fully understood quantum materials \cite{MazinPRL2012,ZaliznyakPNAS2015,BallouPRL1996,TomiyasuPRL2014,HattoriPRB2009,VarmaPRB1997,FengSB2021} where electron itinerancy and correlation are both significant, for which MLWF calculations \cite{PoPRX2018,WaltersNatPhys2009,FoyevtsovaPRB2013,LanePRB2018} may provide important and potentially surprising insights. The conceptual difference between MO and ionic magnetism is profound, and independent of whether the system is metallic or insulating as per band filling. Notably, even if it were an insulator, MnSi would still escape recent high-throughput searches for MO materials \cite{Xu2021,Gao2021}, because both the Mn and the NNET sites share the same Wyckoff label. Finally, we envision magnetic materials with topological bands to be an interesting frontier, where magnetic electrons cannot be described by exponentially localized Wannier orbitals \cite{BradlynNature2017,PoNatComm2017}. In such cases, one might not be able to decompose the magnetic neutron scattering cross section into form factor and spectral functions. A new description embracing the full wave-function structure of the magnetic electrons will be needed.

\begin{acknowledgments}
We are grateful for discussions with Gang Chen, Ji Feng, Jian Kang, Wei Ku, Zhengxin Liu, Zhongyi Lu, Zhida Song, Yuan Wan, Fa Wang, and Yugui Yao. Y.L. acknowledges support by the National Key R\&D Program of China (No. 2018YFA0305602) and the National Natural Science Foundation of China (Nos. 12061131004 and 11888101), and thanks the Tsung-Dao Lee Institute at Shanghai Jiao Tong University for hospitality during the preparation of the manuscript. B.W. acknowledges support by the National Key R\&D Program of China (Nos. 2017YFA0303302 and 2018YFA0305602), the National Natural Science Foundation of China (No. 11921005), and the Shanghai Municipal Science and Technology Major Project (No. 2019SHZDZX01). The work at Brookhaven National Laboratory was supported by Office of Basic Energy Sciences (BES), Division of Materials Sciences and Engineering, U.S. Department of Energy (DOE), under contract DE-SC0012704. Part of this research was performed at the MLF, J-PARC, Japan, under a user program (proposal No. 2019B0085). A portion of this research used resources at Spallation Neutron Source, a DOE Office of Science User Facility operated by the Oak Ridge National Laboratory.

\end{acknowledgments}
\bibliography{main_sm_scibib}
\bibliographystyle{apsrev4-2}
%


\widetext
\begin{center}
\textbf{\large Supplemental Material for ``Magnetic molecular orbitals in {MnSi}''}
\end{center}
\makeatletter
\renewcommand{\theequation}{S\arabic{equation}}
\renewcommand{\thetable}{S\arabic{table}}
\renewcommand{\thefigure}{S\arabic{figure}}
\section{Materials and Methods}
{\bf Sample growth and characterization.}
High-quality single crystals of MnSi were grown by a traveling-solvent floating-zone method. X-ray Laue backscattering from natural crystal surfaces produces sharp diffraction patterns with an approximate two-fold symmetry (Fig.~\ref{Figs2}A), consistent with the cubic space group $P2_13$ (No.~198; $a = 4.556 \rm\AA$). A total of 28 high-quality single crystals (total mass $\approx$ 33 g) were used for our experiments (Fig.~\ref{Figs2}A). The crystals were co-aligned on aluminum plates with a hydrogen-free adhesive, such that reciprocal lattice vectors ($H$,~$L$,~$L$) are in the horizontal plane. The (0,~3,~4) and (0,~4,~3) Bragg reflections differ in their intensities by a factor over 40, indicating that the entire sample is nearly a single domain (Fig.~\ref{Figs2}, B and C). According to a rocking curve measured on the (2,~1,~1) nuclear Bragg reflection, the mosaic spread of the entire sample is about 1.3$^\circ$ (full width at half maximum) (Fig.~\ref{Figs2}D). Magnetometry measurements on a single crystal using a Quantum Design MPMS indicate a magnetic phase transition at 29.5 K and a saturated magnetic moment of 0.4 $\mu_\mathrm{B}/\rm Mn$ (Fig.~\ref{Figs2}, E and F).
\bigbreak
{\bf INS experiment.}
Our INS experiments were performed on the 4SEASONS spectrometer at the MLF, J-PARC, Japan and the SEQUOIA and ARCS spectrometers at SNS, ORNL, USA \cite{4SEASON,SEQUOIA,ARCS}. The 4SEASONS spectrometer has a multiple-${E}_{\rm i}$ capability, such that neutron scattering events with a series of different incident energies are recorded simultaneously. On the 4SEASONS spectrometer, a chopper frequency of 500 Hz was used for ${E}_{\rm i}$ = 22 meV, 29 meV, 40 meV, 57 meV, 90 meV and 162 meV, and measurements were performed at $T$ = 4 K and 40 K. On the SEQUOIA spectrometer, a chopper frequency of 600 Hz was used for ${E}_{\rm i}$ = 90 meV and 150 meV, and measurements were performed at $T$ = 4 K, 40 K and 300 K. On the ARCS spectrometer, chopper frequencies of 360 Hz, 540 Hz and 600 Hz were used for ${E}_{\rm i}$ = 300 meV, 600 meV and 1000 meV, respectively, and measurements were performed at $T$ = 4 K. During the measurements, the sample was rotated about the vertical direction over a 40$^\circ$ range in 0.25$^\circ$ steps on 4SEASONS, and over a 120$^\circ$ range in 1$^\circ$ steps on SEQUOIA and ARCS. Data accumulated at different angles were combined into a four-dimensional data set, which we used the Dave, Utsusemi and Horace software packages to reduce and analyze \cite{Dave,Utsusemi,Horace}. After a careful alignment of the measured data set with the crystallographic coordinate system using all available nuclear Bragg reflections, the entire data set was down-folded into a minimal, physically independent sector of the three-dimensional momentum space using the point-group symmetry of the system. The INS intensities were converted into absolute scattering cross sections by comparing to measurements of vanadium standard samples using exactly the same spectrometer conditions. The resultant cross-sections were further corrected for neutron absorption, which is estimated to cause a minimum of $46\%$ reduction of the scattering intensity based on tabulated data on absorption cross sections for a neutron energy of 90 meV, and with an effective sample thickness of 20 mm. Absorption-corrected absolute cross sections are presented throughout the manuscript.

\bigbreak
{\bf Phonon Calculations.}
The force constants of MnSi were calculated by density functional perturbation theory (DFPT) with Vienna Ab initio Simulation Package (VASP) \cite{Gonze1997,Kresse1996}. Perdew-Burke-Ernzerhof (PBE) type exchange-correlation functionals were adopted, within generalized gradient approximation (GGA) regime \cite{Perdew1996}. The kinetic energy cutoff was set to 400 eV. Integrations over Brillouin zone were performed with Monkhorst-Pack $\bf{Q}$-point grid (equivalent to a $15\times15\times15$ grid). Lattice constants and atomic positions were relaxed until residual forces drop below 0.001 eV/\AA. The relaxed lattice constant was 4.42~\AA, slightly smaller than our experimental value of $4.56\pm 0.01~\rm\AA$ as seen from the neutron diffraction results. The good agreement between the data and calculation shown in Fig.~1A enabled us to distinguish the spin waves from acoustic phonons by inspecting data acquired in different Brillouin zones (BZs) (Fig.~\ref{Figs3}) and to attribute scattering signals to magnetic excitations (1) below 20 meV near the BZ boundary, (2) between 45 and 50 meV, and (3) above 60 meV.

\bigbreak{\bf Maximally Localized Wannier Functions.}
The electronic structure of the near-ferromagnetic phase of MnSi was calculated by VASP within the local density approximation (LDA) \cite{Kresse1996}. An $11\times11\times11$ Monkhorst-Pack grid was used for non-selfconsistent calculations. The cutoff energy for plane wave basis was 550 eV. The magnetic moment was fixed at the experiment value of 0.4 $\mu_{\rm B}$/Mn. We used WANNIER90 to implement further calculations of MLWFs, following the optimization routine proposed by Marzari and Vanderbilt \cite{MOSTOFI2008685,Pizzi2020,Souza2001,MarzariPRB1997}.

\section{Supplementary Text}
{\bf Intensity model for the continuum.}
The central message of our work is that one can experimentally distinguish the magnetic MO's form factor from the ionic one \cite{Shirane2002}. We first verify in Fig.~\ref{Figs11} that the radial $Q$ dependence of the form factor measured on the continuum has an energy-independent momentum structure, namely, the intensity shows a distinct maximum at $Q_{\rm 0} = 2.5 ~{\rm\AA^{-1}}$. To describe the magnetic electron cloud that underlies this behavior, we aim to use a ferromagnetic ``cluster'' of ionic electron clouds \cite{ZaliznyakPNAS2015}, hence the following effective model, to calculate the scattering intensity as a function of momentum transfer,
\begin{equation}
   {I}_{\rm M}({\mathbf{Q}}) = A{\times}
{\left|f(Q)\right|^2}{\times}{\sum_{m}}\left|{\sum_{n}}e^{-i (\mathbf{R}_n\cdot \mathbf{Q})}{\times}e^{-\frac{(\mathbf{R}_n-\mathbf{R}_m)^2}{2{\sigma_1}^2}}\right|^2+B,
\label{EqS1}
\end{equation}
where $A$ is a global normalization factor, $B$ is a constant background, and $f(Q)$ is the (isotropic) atomic form factor of individual $\rm{Mn}$ ions. $\mathbf{R}_m$ denotes the coordinate of the ``correlation center'' in the unit cell which we will explain later, where the summation over $m$ goes through all symmetry-related centers. The summation over $n$ in $\mathbf{R}_n$ goes through all Mn atoms, in order to describe interference on a chosen ferromagnetic cluster centered at $\mathbf{R}_m$. In practice, we only need to consider atoms within a few times of $\lambda = 2\sqrt{2\ln2}\sigma_1$ (size of the cluster) from $\mathbf{R}_m$.

We model the ferromagnetic clusters using three representative choices of the correlation centers: (1) The center of the NNET; there are four equivalent points (the summation over $m$ in Eq.~\ref{EqS1}) in the unit cell. (2) The midpoint between two nearest Mn atoms; there are twelve equivalent points in the unit cell. (3) Mn ions; there are four equivalent points in the unit cell. By adjusting the effective correlation length, we can adjust the size of the magnetic clusters. For instance, when the correlation length is small, under choice (1) the cluster becomes the NNET; under choice (2) the cluster represents a ``diatomic molecule''; under choice (3) the cluster is reduced to the ionic limit. Outputs of the model under the three choices, upon varying the correlation length, are presented in Figs.~\ref{Figs12}–\ref{Figs14}. We find that only choice (1) in the small correlation-length limit can reasonably fit our experiment results. It corresponds to the NNET model that we describe in the main text. In this limit, our model can be simplified as,
\begin{equation}
   {I}_{\rm M}({\mathbf{Q}}) = A {\times}{\left|f(Q)\right|^2}{\times}{\sum_{j=1}^{4}}{\left|F_j(\mathbf{Q})\right|^2}+B,
\label{EqS2}
\end{equation}
where ${F_j(\mathbf{Q})} = {\sum_{{\mathbf{R}}\in{T_j}}}{e^{-i(\mathbf{Q} \cdot \mathbf{R})}}$ summed over Mn positions represents a NNET with four ($j$) possible orientations. The physical meaning of this model is that spins on the three Mn atoms within a NNET are fully coherent and aligned, whereas spins belonging to different NNETs are uncorrelated.

In fact, Eq.~\ref{EqS2} already offers a reasonably global description of intensity variations of the continuum, with respect to  (1) the agreement of simulation and experimental data of the inter-BZ INS intensity at high energy (Fig.~\ref{Figs15}), (2) insensitive dependence on the choice of the nominal valence value which affects $f(Q)$ (Fig.~\ref{Figs16}), and (3) the ratio $\frac{I\rm{(M_1 = (0,~0.5,~0.5))}}{I\rm{(M_2 = (1,~0.5,~0.5))}}$ (Fig.~\ref{Figs17}) consistent with the calculated value of 2.56. But still, the model in Eq.~\ref{EqS2} lacks description of the intra-BZ structures of the continuum. Here, we propose two nearly equally good intra-BZ models, which are to be multiplied with the NNET form factor to describe the INS intensities:

(1) Lattice Lorentzian model: In Figs.~\ref{Figs6} and \ref{Figs7}, the continuum shows a set of broad peaks at $\mathbf{Q} = \mathbf{G} + \mathbf{q}$, positioned periodically in the reciprocal lattice. When considering the inter-NNET correlation in real space and using the normalized lattice Lorentzian function for the continuum peak \cite{ZaliznyakPNAS2015}, we obtain
\begin{equation}
   {I}_{\rm M}({\mathbf{Q}}) = A {\times}{\left|f(Q)\right|^2}\times\left(\sum_{j=1}^{4}{\left|F_j(\mathbf{Q})\right|^2}\right){\times}
\mathrm{LL}({\mathbf{Q}})+B,
\label{EqS3}
\end{equation}
\begin{equation}
\begin{split}
   \mathrm{LL}({\mathbf{Q}}) =
\mathrm{\frac{1}{2}(\frac {sinh(\xi^{-1})}{cosh(\xi^{-1})-cos(2\pi(\it{H-h_0}))}+\frac {sinh(\xi^{-1})}{cosh(\xi^{-1})-cos(2\pi(\it{H+h_0}))})}\\
{\times}\mathrm{\frac{1}{2}(\frac {sinh(\xi^{-1})}{cosh(\xi^{-1})-cos(2\pi(\it{K-k_0}))}+\frac {sinh(\xi^{-1})}{cosh(\xi^{-1})-cos(2\pi(\it{K+k_0}))})}\\
{\times}\mathrm{\frac{1}{2}(\frac {sinh(\xi^{-1})}{cosh(\xi^{-1})-cos(2\pi(\it{L-l_0}))}+\frac {sinh(\xi^{-1})}{cosh(\xi^{-1})-cos(2\pi(\it{L+l_0}))})},
\end{split}
\label{EqS4}
\end{equation}
where $\mathrm{LL}({\mathbf{Q}})$ is the normalized lattice Lorentzian function, $\xi = \xi_a = \xi_b = \xi_c$ is the correlation length in real space, $h_0 = k_0 =l_0$ is the continuum peak position. The best-fit parameters are $h_0 = k_0 = l_0 = 0.5$ r.l.u. and $\xi = 0.35$ l.u.

(2) Spherical shell model: Since the continuum is located on a thick and hollow spherical shell within the BZ (Fig.~\ref{Fig2}C), by modifying Eq.~\ref{EqS2} with a spherical shell function, we obtain
\begin{equation}
   {I}_{\rm M}({\mathbf{Q}}) = A {\times}\left|f(Q)\right|^2\times\left(\sum_{j=1}^{4}{\left|F_j(\mathbf{Q})\right|^2}\right){\times}\mathrm{Shell}({\mathbf{Q}})+B,
\label{EqS5}
\end{equation}
\begin{equation}
   \mathrm{Shell}({\mathbf{Q}}) = \int_{\Omega={\{\left|{\mathbf{Q}^{\prime}-\mathbf{G}}\right| = R_{0}}\}}e^{-({\frac{\mathbf{Q} -\mathbf{Q}^{\prime} } {\sqrt{2}\sigma_0}})^2}{\rm d}\mathbf{Q}^{\prime},
\label{EqS6}
\end{equation}
where $\mathrm{Shell}({\mathbf{Q}})$ is the spherical shell function, $\mathbf{G}$ is the nearest integer Bragg point from $\mathbf{Q}^{\prime}$, and $R_0$ and $\sigma_0$ describe the radius and Gaussian thickness of spherical shells, respectively. The best-fit parameters are $R_0 = 0.76~\rm\AA^{-1}$ and $\sigma_0 = 0.39~\rm\AA^{-1}$.

\bigbreak
{\bf Spin-wave calculation and spectral weight of the continuum.}
To understand the observed (nearly) ferromagnetic spin waves, we fit the spin-wave spectra with the SpinW program \cite{Toth_2015}. The magnetic Heisenberg Hamiltonian reads
\begin{equation}
   {H} = -{\sum_{i<j}}{J_{ij}}{S_i}{\cdot {S_j}},
\label{EqS7}
\end{equation}
where $J_{ij}$ is the magnetic exchange coupling of the spin $S_i$ and $S_j$. Assuming that the nearest-neighbor ferromagnetic exchange coupling $J_1$ ($>0$) dominates, our best-fit values are $J_1$ = 21.8 meV and $0.7\pm 0.15$ $\mu_{\rm B}$/Mn, with the Lande g-factor set equal to 2 and the magnetic form factor of $\rm{Mn}^{2+}$ ion, as shown in Fig.~\ref{Figs19}. Here, for the calculation of the ``acoustic'' spin-wave branch, it is still a good approximation to consider all the magnetic moments to reside on the Mn ions, because spin waves in this limit are nearly uniform motion of the magnetization density, and there is no difference from choosing the NNETs as the fundamental units. The approximation gradually breaks down as the wavelength approaches the spatial extent of the NNETs (see text).

Figure~\ref{Figs20} shows the imaginary part of the dynamic magnetic susceptibility of the continuum (momentum-integrated in (2, 1, 0) BZ) \cite{Xuguangyong}, based on the absolute scattering cross sections and after accounting for the NNET form factor. The energy-integrated spectral weight from 45 meV to 260 meV in Fig.~\ref{Figs20} is $1.1 \pm 0.2$ $\mu_\mathrm{B}^2/\mathrm{Mn}$. Due to the contamination of phonons below 45 meV, here we use a damped harmonic oscillator function (DHO) to fit $\chi^{\prime\prime}(E)$ and extrapolate the data to zero and much higher energies ($\sim$ 1 eV)\cite{ZaliznyakPNAS2015}, which amounts to a total spectral weight $2.2\pm 0.4$ $\mu_\mathrm{B}^2/\mathrm{Mn}$.

\bigbreak
{\bf EBR-Restricted Maximally Localized Wannier Function Calculation.}
A suitable tight-binding (TB) basis is essential for properly describing low-energy physics such as magnetism in crystalline materials. While it is often taken for granted that atomic orbitals form such a basis in magnetic compounds, this assumption is problematic when orbital hybridization is significant. In such cases, obtaining a physically trackable and mathematically rigorous TB model may be a nontrivial task. Rather than relying on intuition and inspection of DFT-calculated bands, which do provide valuable insights in the presence of certain characteristic features in the band structure such as an isolated flat band \cite{FoyevtsovaPRB2013}, here we adopt the concepts of band representation (BR) and elementary band representation (EBR), which have recently been developed in the context of topological quantum chemistry \cite{BradlynNature2017}, to guide our determination of suitable TB bases.

A BR is a collection of energy bands that are energetically isolated from the other bands. Generically, it may or may not be possible to reproduce a given BR using TB models based on local orbitals that respect the crystallographic symmetries (and when this is not possible, the BR is topological \cite{BradlynNature2017,PoNatComm2017}). An EBR, in contrast, is defined by the formation of energy bands starting from a minimal set of local orbitals consistent with the symmetry -- the orbitals must be an irrep of a maximal Wyckoff site's point group \cite{BradlynNature2017,PoNatComm2017,Elcoroks5574}. In other words, the simplest form of BR that can possibly be obtained with a TB model is an EBR. More generally, BRs that can be described by TB models must be decomposable into EBRs, whereas the process of EBR identification and BR decomposition (into EBR) can be facilitated by symmetry indicators \cite{BradlynNature2017,PoNatComm2017}. An important example (in non-topological materials) is a non-degenerate and isolated (flat) band, which is always an EBR. Such band's existence near the Fermi level immediately calls for a ``stand-alone'' TB description of the low-energy physics, yet this simple intuition is available only when there is at least one fixed point in the primitive cell that remains unchanged under all space-group operations.

Importantly, an EBR does not have to be, and is often not, a single band. Yet it remains to be a \textit{defining} property of a minimal TB model because of symmetry. In our present case, since the space group of MnSi is non-symmorphic, all EBRs (a total of four) contain either 4 or 8 bands. Our strategy here is hence to first identify bands near the Fermi energy as EBRs (or decomposable BRs), and then to find Wannier orbitals for each of the EBRs and BRs as a whole. To fix the gauge of the Wannier orbitals, we adopt the maximal localization criterion \cite{MarzariRMP2012}. We call this method ``EBR-restricted MLWF calculation.'' The strongest constraint is set by the identification of EBRs and BRs, which ensures that the obtained TB model is a low-energy minimal model. Meanwhile, the maximal localization requirement not only satisfies the tight-binding philosophy, but also makes the computation of magnetic neutron scattering cross section (as a function of momentum) most conveniently separable into intra- and inter-BZ parts, the latter of which is our form factor, and the next Section provides a mathematical proof of it.

We have calculated the irreps of the 8 conduction bands at high-symmetry points of the BZ using the open-source program irvsp \cite{GAO2021107760,Vergniory2019}. The result is listed in Table \ref{band representation}. By comparing to the EBR table for space group No.~198 on the Bilbao Crystallographic Server
({https://www.cryst.ehu.es/cgi-bin/cryst/programs/bandrep.pl})
\cite{Elcoroks5574}, it is clear that these bands constitute the EBR labeled as $E\uparrow G$, where $E$ is a two-dimensional irrep of the site symmetry group $C_3$ of Wyckoff site $4a$, and $E\uparrow G$ indicates induced representation on space group $G$ (of MnSi). Similarly, we find that the next 12 bands deeper below $E_\mathrm{F}$ are an $(A_1+ E)\uparrow G$ composite BR. These two EBR/BRs are entangled only slightly with each other, making it relatively straightforward to obtain high-quality MLWFs for each of them.

The minimal TB model for low-energy physics is obtained by using the Bloch states in the EBR $E\uparrow G$ as input. The model is expected to involve the Wyckoff $4a$ site, but because $4a$ is a ``variable'' site whose coordinate contains a continuous parameter (Fig.~\ref{Figs1}), the $4a$ label alone does not yet allow us to pin down the magnetization clouds (namely magnetic Wannier functions). The gauge freedom of Wannier functions is fixed by the maximally localized criterion. After minimizing the spreads of the Wannier functions, the Wannier centers of the EBR $E\uparrow G$ turn out to be very close to the centers of the NNETs (Fig.~\ref{Fig4}B, Fig.~\ref{Figs21}C and Fig.~\ref{Figs22}, C1-C3), which deviate significantly from the Mn atomic sites and signals formation of molecular orbitals (MOs).  It means that the $E$ in $E\uparrow G$ comes from the molecular sites $M_{i=1,2,3,4}$ (i.e centers of the NNETs, the corresponding local $z$ axes for $i=1,2,3,4$ are $[\bar1 1 \bar1],[1 \bar1 \bar1],[\bar1 \bar1 1],[1 1 1]$, and the local $x$ axes are respectively $[1 1 0],[\bar1 \bar1 0],[1 \bar1 0],[\bar1 1 0]$) rather than the Mn (or Si) atomic sites, even though they all share the same Wyckoff label $4a$.

While surprising at first sight, these NNET-centered MLWFs are a robust result. To see it, we compare them to MLWFs calculated under Mn-site constraint. As both sets of MLWFs are constructed from the same EBR's Bloch states, the resultant TB models reproduce the band dispersion comparably well (Fig.~\ref{Figs21}A). However, the constrained MLWFs are more extended, seen from their 36\% larger spread (15.4~$\mathrm{\AA}^2$, compared to 11.3~$\mathrm{\AA}^2$ of the NNET-centered MLWFs) and considerably wider contour surface enclosing half of the electron weight (Fig.~\ref{Figs21}, C and D). This difference is significant, because a Wannier orbital's effectiveness in describing the INS cross section explicitly depends on its localization property (see the proof in next section). The constrained MLWFs are also dissimilar in shape to the AOs of Mn, which should not be the case if an ionic picture were effective. Meanwhile, the MLWFs are calculated to be the AOs of Mn ($e$ orbitals \cite{ShanavasPRB2016}) only when a much greater number of Bloch states -- a total of 32 bands near $E_\mathrm{F}$ -- are taken as input. Despite being localized (Fig.~\ref{Figs21}E), the AOs fail to account for the low-energy physics (Fig.~\ref{Figs21}B): the TB model clearly deviates from the magnetic EBR when hybridizations with other AOs (primarily $p$ of Si) are neglected. The hybridizations push the AOs' partial density of states (PDOS) into bands both above and far below $E_\mathrm{F}$ \cite{JeongPRB2004}, indicating covalency of the system. In Figs.~\ref{Figs22} and \ref{Figs23}, we further show that \textit{all} electronic states within $\sim6$ eV below $E_\mathrm{F}$ are MOs, among which the PDOS near $E_\mathrm{F}$ is exclusively contributed by the MOs on the NNETs. With all the supporting evidence, we conclude that the experimentally revealed magnetic MOs are fully consistent with our EBR-restricted MLWF calculation.

The MLWFs on a given molecular site form a two-dimensional irrep $E$ of the site symmetry group \cite{Smirnov,Sakuma} $G_{M_{i}}$ (isomorphic to $C_{3,\langle111\rangle}$), and we label them as $W_{M_{i},\mu=a,b}$ (Fig.~\ref{Figs22}, C1 and C2, the lattice-cell index $R$ is omitted for simplicity; note that for the convenience of plotting, we have chosen a real basis, such that each of the wave functions is off-center from $M_i$ and slightly less localized than the complex MLWFs). Band interpolation using these MOs as the TB basis is highly satisfactory (Fig.~\ref{Fig4}A, Fig.~\ref{Figs21}A, Fig.~\ref{Figs22}A and Fig.~\ref{Figs23}A). The intrinsic property of the MLWFs can be characterized by the $C_3$-invariant density function $n_{M_{i}}\equiv\frac{1}{2}\sum_{\mu=a,b}\left|W_{M_{i},\mu}\right|^{2}$, which is visualised in Fig.~\ref{Figs22}, C1-C3.

The above MOs $W_{M_{i},\mu}$ of the magnetic EBR (Fig.~\ref{Figs22}, C1 and C2) form a ``stand-alone'' subspace in the electronic structure. To see this, we need to show that an electron in these MOs can hardly hop into orbitals (whether or not in the same lattice cell) outside the subspace. We use a two-orbital hopping toy model to set the stage: for a $2\times2$ matrix Hamiltonian, in which hopping $t$ is the off-diagonal elements, the energy-level difference measured from the average is given by $\sqrt{(\delta E/2)^2+t^2}$, where $\delta E \equiv E_1 - E_2$ is the energy difference without the hopping. Using such a pairwise description for the Wannier orbitals, the significance of hopping can be quantified by a collection of dimensionless numbers $\eta_{ij}(R)=\left|E_i-E_j\right|/\sqrt{\left|E_i-E_j\right|^2+4t_{ij}(R)^2}$, where $t_{ij}(R) = \left\langle W_{i}(0)\left|H\right|W_{j}(R)\right\rangle$ is the hopping strength between the $i$-th and the $j$-th Wannier orbitals $R$ cells apart, and $E_{i} = \left\langle W_{i}(0)\left|H\right|W_{i}(0)\right\rangle$ is the on-site energy. By definition, $\eta$ is between 0 and 1, and the two limits are hopping- and gap-dominated, respectively. If the MOs of the magnetic EBR indeed form a ``stand-alone'' subspace, an electron in any one of the MOs should be able to hop into at least some other MOs in the same EBR, giving rise to an $\eta$ close to 0 (as the $E_i$'s are the same), but it cannot hop into any of the Wannier orbitals outside the EBR, so that those $\eta$ should be close to 1. The last two columns of Table~\ref{MOhop} show exactly this [for $W_{M_{4},a}$, displayed in Fig.~\ref{Figs22}C1, chosen without loss of generality]. The result is nothing but an alternative way to see that the TB bands formed by the MOs are well-separated from the other bands, as visualized in Fig.~\ref{Figs22}A.

Table~\ref{MOhop} further enables us to show that our EBR-restricted Wannier orbital computation is mathematically more rigorous and trackable than a conventional method of using atomic orbitals (AOs) to construct MOs, as has been done for Na$_2$IrO$_3$ \cite{FoyevtsovaPRB2013}. To take the latter approach, one would need to first treat the symmetry properties of the AOs carefully \cite{Sakuma}, and then use physical intuition (such as with quantum chemistry) to construct MOs that can only ``hop into themselves'' by virtue of hopping cancellation between the constituent AOs. What we explain below and show in Table~\ref{MOhop} is that, for MnSi, taking such an approach is a formidable (and, practically incomprehensible) task due to the system's significant electron itinerancy.

To ensure that we start with an adequate mathematical description of the task, we have computed 32 atomically centered maximally localized Wannier functions [Mn $3d = a_{3z^2-r^2} + (e_{xy},\,e_{x^2-y^2})+(e_{xz},\,e_{yz})$ and Si $3p = a_z + (e_x,\,e_y)$ orbitals, $z$ denotes the local $C_3$ axis; there are 4 atoms of each type in the primitive cell]. They respect the atomic-site symmetries and allow us to reproduce the full band structure ranging from $\sim6.5$ eV below to $\sim6.5$ eV above the Fermi level. The 8 AOs that have the largest contribution to the magnetic EBR are displayed in Fig.~\ref{Figs21}. Limited by the $11\times11\times11$ $\mathbf{k}$-grid used in our non-selfconsistent calculations, we are then able to place these AOs on an equal grid of the lattice-cell index $R$, forming a basis containing a total of 42592 AOs. A chosen MO from the magnetic EBR (1 out of 8) situated in the home cell can then be projected onto the AO basis, in order to reveal how the MO can be constructed from the AOs. We note that the numerical accuracy of our MOs is very high -- the normalization of the 8 MOs in the home cell is $1-\delta_\mathrm{N}$ with the largest $\left|\delta_\mathrm{N}\right|=0.002$, and the orthogonality between them is $\delta_\mathrm{O}$ with the largest $\left|\delta_\mathrm{O}\right|=6\times 10^{-6}$. The AOs are similarly accurate.

The main conclusions following from the data summarized in Table~\ref{MOhop} are two-fold: First, due to the spatial extent of the MOs which reflects electron itinerancy, the projection of an MO onto AOs up to a cut-off distance from the MO center only becomes ``complete'' when the cut-off distance is extremely large (12 $\mathrm{\AA}$, over twice the lattice constant, in order to lose no more than 1\% of weight). Second, while the hopping into nearby MOs in the same EBR is always strong (``$\eta_\mathrm{min}$ within EBR'' in Table~\ref{MOhop}), hopping outside the EBR cannot be completely eliminated even up to such a large cut-off distance, although a converging trend is seen as the distance increases. Indeed, to obtain a decent AO weight and elimination of unwanted hopping, one would need to include as many as 147 AOs up to a distance of 4.6 $\mathrm{\AA}$, whereas including only 61 AOs up to a distance of 3.6 $\mathrm{\AA}$ already makes the result considerably worse.

To provide some concrete numbers, the MO $W_{M_{4},a}$ is a linear combination AOs ``mainly'' contributed by two nearby Mn atoms [see Fig.~\ref{Figs22}(C1) for their relative locations]: $W_{M_{4},a}=-0.41(\text{Mn}_1)_{x^2-y^2}-0.35(\text{Mn}_1)_{3z^2-r^2} +0.25(\text{Mn}_2)_{3z^2-r^2}+0.23(\text{Mn}_1)_{yz}+0.22(\text{Mn}_2)_{yz}+0.20(\text{Mn}_2)_{xy} -0.17(\text{Mn}_2)_{x^2-y^2}+\cdots$, where each of the atoms has its own $z$ axis defined by the local $C_3$ symmetry and different from the $z$ axis of $M_4$. However, summing up the above-listed (squared) weight of the AOs produces only 52\% of the MO, the rest of which is contributed by many more AOs at further distances from the MO center. This clearly shows the difficulty of manually constructing MOs from AOs in the presence of strong electron itinerancy. Importantly, the multi-orbital superposition from the same Mn atoms indicates that the MOs are subject to strong Hund's rule coupling, which promotes the ferromagnetism in the seemingly strongly dispersing magnetic EBR.

The wave function's sign reversal between the two lead-contribution atoms [Fig.~\ref{Figs22}, C1 and C2] indicates that the MO is of anti-bonding nature: the electron density is small around the Mn-Mn bond center, and the NNET's magnetization cloud has a ``hollow'' internal structure, which has been directly confirmed by our INS experiment. We have used the crystal orbital Hamiltonian population (COHP) method \cite{Dronskowski1993,KomlevaPRB2020} to verify the anti-bonding character near the Fermi energy (Fig.~\ref{Figs23}B). Since formation of covalent bonds is energetically motivated by the occupancy of bonding rather than anti-bonding states, for completeness, we have performed MLWF calculations for the aforementioned 12-band BR deeper below $E_\mathrm{F}$. To our satisfaction, this BR is indeed formed by two bonding-typed MOs with high electron densities between nearby Mn and Si atoms (Fig.~\ref{Figs22}). We thus conclude that a thorough microscopic explanation for the formation of magnetic MOs has been obtained for MnSi.

\newpage
{\bf Cross sections of Stoner excitations from a spin-polarized EBR.}
Here we derive the magnetic neutron scattering cross sections of Stoner excitations. It is concluded that a form-factor-like quantity associated with the maximally-localized Wannier functions (MLWFs) of the magnetic bands provides a good approximation for the wide-range momentum dependence of the cross section, given that the relevant bands all belong to a common elementary band representation (EBR) in the unpolarized state. As an EBR can be described by a minimal tight-binding model of essentially a single orbital (and its symmetry-related counterparts) on the crystal lattice, once the orbital is determined as the MLWFs, a spin-flip operator's matrix elements between Bloch states can be conveniently computed from the MLWFs' electron densities.

The spin operator $\hat{S}^{\alpha}$ ($\alpha=x,y,z$) can be written as
\[
\hat{S}^{\alpha}=\hat{\Psi}^{\dagger}\left(r\right)S^{\alpha}\hat{\Psi}\left(r\right)=\sum_{\sigma\sigma^{\prime}}\hat{\psi}_{\sigma}^{\dagger}\left(r\right)S_{\sigma\sigma^{\prime}}^{\alpha}\hat{\psi}_{\sigma^{\prime}}\left(r\right),
\]
where $S^{\alpha}$ are the Pauli matrices and $\hat{\Psi}^{\dagger}\left(r\right)=\left[\begin{array}{cc}
\hat{\psi}_{\uparrow}^{\dagger}\left(r\right) & \hat{\psi}_{\downarrow}^{\dagger}\left(r\right)\end{array}\right]$ is the creation operator of spinor wave functions. $\hat{\psi}_{\sigma}^{\dagger}\left(r\right)$
can be expressed \cite{Cooke1973} by the creation operator $\hat{C}_{n,k,\sigma}^{\dagger}$ of single-particle
Bloch states $\psi_{n,k,\sigma}\left(r\right)$ ($n$: band index; $k$: wave vector; hereafter, all position and momentum variables are vectors)
\[
\hat{\psi}_{\sigma}^{\dagger}\left(r\right)=\frac{1}{N}\sum_{nk}\psi_{n,k,\sigma}^{*}\left(r\right)\hat{C}_{n,k,\sigma}^{\dagger},
\]
where $N$ is the number of mesh points in the Brillouin zone and the normalization
condition is $\left\langle \psi_{n,k}\mid\psi_{m,k^{\prime}}\right\rangle =N\delta_{nm}\delta_{kk^{\prime}}$.
In the basis of $\psi_{n,k,\sigma}\left(r\right)$, the spin operator
is given by
\[
\hat{S}^{\alpha}=\frac{1}{N^{2}}\sum_{k,n,m,k^{\prime},\sigma,\sigma^{\prime}}\psi_{n,k,\sigma}^{*}\left(r\right)S_{\sigma\sigma^{\prime}}^{\alpha}\psi_{m,k^{\prime},\sigma^{\prime}}\left(r\right)\hat{C}_{n,k,\sigma}^{\dagger}\hat{C}_{m,k^{\prime},\sigma^{\prime}},
\]
and its Fourier transform is
\begin{align*}
\hat{S}_{q}^{\alpha} & =\int d^{3}re^{-\mathrm{i}qr}\hat{S}^{\alpha}=\int d^{3}re^{-\mathrm{i}qr}\frac{1}{N^{2}}\sum_{k,n,m,k^{\prime},\sigma,\sigma^{\prime}}\psi_{n,k,\sigma}^{*}\left(r\right)S_{\sigma\sigma^{\prime}}^{\alpha}\psi_{m,k^{\prime},\sigma^{\prime}}\left(r\right)\hat{C}_{n,k,\sigma}^{\dagger}\hat{C}_{m,k^{\prime},\sigma^{\prime}}\\
 & =\frac{1}{N^{2}}\int d^{3}re^{-\mathrm{i}qr}\sum_{k,n,m,\sigma,\sigma^{\prime}}\psi_{n,k,\sigma}^{*}\left(r\right)S_{\sigma\sigma^{\prime}}^{\alpha}\psi_{m,k+q,\sigma^{\prime}}\left(r\right)\hat{C}_{n,k,\sigma}^{\dagger}\hat{C}_{m,k+q,\sigma^{\prime}}.
\end{align*}
The transverse components, which produce spin flips, are
\[
\hat{S}_{q}^{-}=\int d^{3}re^{-\mathrm{i}qr}\hat{S}^{-}=\frac{1}{N^{2}}\int d^{3}re^{-\mathrm{i}qr}\sum_{k,n,m}\psi_{n,k,\downarrow}^{*}\left(r\right)\psi_{m,k+q,\uparrow}\left(r\right)\hat{C}_{n,k,\downarrow}^{\dagger}\hat{C}_{m,k+q,\uparrow},
\]
\[
\hat{S}_{q}^{+}=\int d^{3}re^{-\mathrm{i}qr}\hat{S}^{+}=\frac{1}{N^{2}}\int d^{3}re^{-\mathrm{i}qr}\sum_{k,n,m}\psi_{n,k,\uparrow}^{*}\left(r\right)\psi_{m,k+q,\downarrow}\left(r\right)\hat{C}_{n,k,\uparrow}^{\dagger}\hat{C}_{m,k+q,\downarrow},
\]
where we use the matrix representation $S^{-}=\left[\begin{array}{cc}
0 & 0\\
1 & 0
\end{array}\right]$ and $S^{+}=\left[\begin{array}{cc}
0 & 1\\
0 & 0
\end{array}\right]$. $\hat{S}_{q}^{-}$ and $\hat{S}_{-q}^{+}$ are Hermitian
conjugates.

It is useful to transform from Bloch functions to Wannier
functions $W_{l_{d},l_{s},\sigma}\left(r-R\right)\equiv W_{l_{d},\sigma}\left(r-l_{s}-R\right)$, where $R$ is
a lattice vector, $l_{s}$ is a Wyckoff site ($l_{s}=1,\cdots,n_{s}$
with site multiplicity $n_{s}$) in the primitive
cell, and $l_{d}=1,\cdots,d$ is the index within a $d$-dimensional
irrep of the Wyckoff site. Together, the local orbitals induce the considered bands,
and when the Wyckoff site is a maximal Wyckoff site, the induced bands form
an EBR \cite{BradlynNature2017}.
In this case, we emphasize that although
the number of bands $(n_{s} d)$ can generally be greater than one, the bands
cannot be further separated into smaller sets of bands because of band
degeneracy and compatibility relationships enforced by symmetry.
We denote $l=(l_{d},l_{s})$ to simplify the notation. According to
the transformation
\[
\psi_{n,k,\sigma}\left(r\right)=\sum_{l,R}e^{\mathrm{i}kR}W_{l,\sigma}\left(r-R\right)U_{l,n,\sigma}^{\dagger}\left(k\right),
\]
\[
\hat{C}_{n,k,\sigma}^{\dagger}=\sum_{l,R}e^{\mathrm{i}kR}U_{l,n,\sigma}^{\dagger}\left(k\right)\hat{T}_{l,R,\sigma}^{\dagger},
\]
where $\hat{T}_{l,R,\sigma}^{\dagger}$ is the creation operator of
Wannier function $\left|W_{l,R,\sigma}\right\rangle $,  we have
\begin{align*}
\hat{S}_{q}^{-} & =\frac{1}{N^{2}}\int d^{3}re^{-\mathrm{i}qr}\sum_{k,n,m}\psi_{n,k,\downarrow}^{*}\left(r\right)\psi_{m,k+q,\uparrow}\left(r\right)\hat{C}_{n,k,\downarrow}^{\dagger}\hat{C}_{m,k+q,\uparrow}\\
 & =\frac{1}{N^{2}}\sum_{k,n,m}\sum_{l,l^{\prime},R,R^{\prime}}\int e^{-\mathrm{i}kR}W_{l\downarrow}^{*}\left(r-R\right)U_{nl\downarrow}\left(k\right)e^{-\mathrm{i}qr}e^{\mathrm{i}\left(k+q\right)R^{\prime}}W_{l^{\prime}\uparrow}\left(r-R^{\prime}\right)U_{l^{\prime}m\uparrow}^{\dagger}\left(k+q\right)d^{3}r\hat{C}_{n,k,\downarrow}^{\dagger}\hat{C}_{m,k+q\uparrow}\\
 & =\frac{1}{N^{2}}\sum_{k,n,m}\sum_{l,l^{\prime},R,R^{\prime}}\int e^{-\mathrm{i}k\left(R-R^{\prime}\right)}W_{l\downarrow}^{*}\left(r-R\right)U_{nl\downarrow}\left(k\right)e^{-\mathrm{i}qr}e^{\mathrm{i}qR^{\prime}}W_{l^{\prime}\uparrow}\left(r-R^{\prime}\right)U_{l^{\prime}m\uparrow}^{\dagger}\left(k+q\right)d^{3}r\hat{C}_{n,k,\downarrow}^{\dagger}\hat{C}_{m,k+q\uparrow}\\
 & =\frac{1}{N^{2}}\sum_{k,n,m}\sum_{l,l^{\prime},R,R^{\prime}}e^{-\mathrm{i}k\left(R-R^{\prime}\right)}U_{nl\downarrow}\left(k\right)U_{l^{\prime}m\uparrow}^{\dagger}\left(k+q\right)\int W_{l\downarrow}^{*}\left(r-\left(R-R^{\prime}\right)\right)e^{-\mathrm{i}qr}W_{l^{\prime}\uparrow}\left(r\right)d^{3}r\hat{C}_{n,k,\downarrow}^{\dagger}\hat{C}_{m,k+q\uparrow}\\
 & =\frac{1}{N^{2}}\sum_{k,n,m}\sum_{l,l^{\prime},R,R^{\prime}}e^{-\mathrm{i}k\left(R-R^{\prime}\right)}U_{nl\downarrow}\left(k\right)U_{l^{\prime}m\uparrow}^{\dagger}\left(k+q\right)\tilde{\rho}_{ll^{\prime},R-R^{\prime}}^{W}\left(q\right)\hat{C}_{n,k,\downarrow}^{\dagger}\hat{C}_{m,k+q\uparrow}\\
 & =\frac{1}{N}\sum_{k,n,m}\sum_{l,l^{\prime},R}e^{-\mathrm{i}kR}U_{nl\downarrow}\left(k\right)U_{l^{\prime}m\uparrow}^{\dagger}\left(k+q\right)\tilde{\rho}_{ll^{\prime},R}^{W}\left(q\right)\hat{C}_{n,k,\downarrow}^{\dagger}\hat{C}_{m,k+q\uparrow}\\
 & =\frac{1}{N}\sum_{k,n,m}\sum_{l,l^{\prime},R}e^{-\mathrm{i}kR}U_{nl\downarrow}\left(k\right)U_{l^{\prime}m\uparrow}^{\dagger}\left(k+q\right)\tilde{\rho}_{ll^{\prime},R}^{W}\left(q\right)\cdot\sum_{l_{1}l_{2},R_{1}R_{2}}e^{\mathrm{i}kR_{1}-\mathrm{i}\left(k+q\right)R_{2}}U_{l_{1}n\downarrow}^{\dagger}\left(k\right)U_{ml_{2}\uparrow}\left(k+q\right)\hat{T}_{l_{1}R_{1}\downarrow}^{\dagger}\hat{T}_{l_{2}R_{2}\uparrow}\\
 & =\frac{1}{N}\sum_{k}\sum_{l,l^{\prime},R}e^{-\mathrm{i}kR}\tilde{\rho}_{ll^{\prime},R}^{W}\left(q\right)\sum_{R_{1}R_{2}}e^{\mathrm{i}kR_{1}-\mathrm{i}\left(k+q\right)R_{2}}\hat{T}_{lR_{1}\downarrow}^{\dagger}\hat{T}_{l^{\prime}R_{2}\uparrow}\\
 & =\sum_{R_{1}R_{2}}\sum_{l,l^{\prime}}\tilde{\rho}_{ll^{\prime},R_{1}-R_{2}}^{W}\left(q\right)e^{-\mathrm{i}qR_{2}}\hat{T}_{lR_{1}\downarrow}^{\dagger}\hat{T}_{l^{\prime}R_{2}\uparrow},
\end{align*}
where $\rho_{ll^{\prime},R}^{W}\left(r\right)\equiv W_{l\downarrow}^{*}\left(r-R\right)W_{l^{\prime}\uparrow}\left(r\right)$
is the density matrix of the Wannier functions and its Fourier transform is
\[
\tilde{\rho}_{ll^{\prime},R}^{W}\left(q\right)\equiv\int\rho_{ll^{\prime},R}^{W}\left(r\right)e^{-\mathrm{i}qr}d^{3}r=\int W_{l\downarrow}^{*}\left(r-R\right)W_{l^{\prime}\uparrow}\left(r\right)e^{-\mathrm{i}qr}d^{3}r.
\]
We further transform the operator basis to $k$ space: $\hat{T}_{lR_{1}\downarrow}^{\dagger}=\frac{1}{N}\sum_{k}e^{-\mathrm{i}kR_{1}}\hat{T}_{lk\downarrow}^{\dagger}$. The states created by $\hat{T}_{lk\sigma}^{\dagger}$ are the so-called generalized
Bloch states $\left|\tilde{\psi}_{l,k,\sigma}\right\rangle \equiv \sum_{n}U_{nl\sigma}\left(k\right)\left|\psi_{n,k,\sigma}\right\rangle $,
which are labeled by the local index $l$ rather than the band index $n$, and which are typically not eigenstates of the Hamiltonian any more (see Eq.(8) in \cite{MarzariRMP2012}). The energy of $\left|\tilde{\psi}_{l,k,\sigma}\right\rangle$ is give by
\begin{align*}
E_{l,k,\sigma}=\left\langle 0\left|\hat{T}_{lk\sigma}\hat{H}\hat{T}_{lk\sigma}^{\dagger}\right|0\right\rangle  & =\sum_{R_{1}R_{2}}e^{\mathrm{i}k\left(R_{1}-R_{2}\right)}\left\langle W_{l,\sigma}\left(r-R_{2}\right)\left|\hat{H}\right|W_{l,\sigma}\left(r-R_{1}\right)\right\rangle \\
 & =\sum_{R}e^{\mathrm{i}kR}\left\langle W_{l,\sigma}\left(r\right)\left|\hat{H}\right|W_{l,\sigma}\left(r-R\right)\right\rangle.
\end{align*}

In this new basis, a spin-flip operator becomes
\begin{align*}
\hat{S}_{q}^{-} & =\sum_{R_{1}R_{2}}\sum_{l,l^{\prime}}\tilde{\rho}_{ll^{\prime},R_{1}-R_{2}}^{W}\left(q\right)e^{-\mathrm{i}qR_{2}}\hat{T}_{lR_{1}\downarrow}^{\dagger}\hat{T}_{l^{\prime}R_{2}\uparrow}\\
 & =\sum_{R_{1}R_{2}}\sum_{l,l^{\prime}}\int d^{3}re^{-\mathrm{i}qr}W_{l\downarrow}^{*}\left(r-R_{1}\right)W_{l^{\prime}\uparrow}\left(r-R_{2}\right)\frac{1}{N^{2}}\sum_{kk^{\prime}}e^{-\mathrm{i}kR_{1}+\mathrm{i}k^{\prime}R_{2}}\hat{T}_{lk\downarrow}^{\dagger}\hat{T}_{l^{\prime}k^{\prime}\uparrow}\\
 & =\frac{1}{N}\sum_{k}\sum_{R_{1}R_{2}}\sum_{l,l^{\prime}}\int d^{3}re^{-\mathrm{i}qr}W_{l\downarrow}^{*}\left(r-R_{1}\right)W_{l^{\prime}\uparrow}\left(r-R_{2}\right)e^{-\mathrm{i}kR_{1}+\mathrm{i}\left(k+q\right)R_{2}}\hat{T}_{lk\downarrow}^{\dagger}\hat{T}_{l^{\prime}k+q\uparrow}\\
 & =\frac{1}{N}\sum_{k}\sum_{R_{1}R_{2}}\sum_{l,l^{\prime}}\tilde{\rho}_{ll^{\prime},R_{1}-R_{2}}^{W}\left(q\right)e^{-\mathrm{i}kR_{1}+\mathrm{i}kR_{2}}\hat{T}_{lk\downarrow}^{\dagger}\hat{T}_{l^{\prime}k+q\uparrow}\\
 & =\sum_{k}\sum_{l,l^{\prime}}\left(\sum_{R}\tilde{\rho}_{ll^{\prime},R}^{W}\left(q\right)e^{-\mathrm{i}kR}\right)\hat{T}_{lk\downarrow}^{\dagger}\hat{T}_{l^{\prime}k+q\uparrow}.
\end{align*}
Assuming that the MLWFs are sufficiently localized, the leading
term is contributed by the $R=0$ condition:
\[
\hat{S}_{q}^{-}\approx\sum_{k}\sum_{l,l^{\prime}}\int W_{l\downarrow}^{*}\left(r\right)W_{l^{\prime}\uparrow}\left(r\right)e^{-\mathrm{i}qr}d^{3}r\hat{T}_{lk\downarrow}^{\dagger}\hat{T}_{l^{\prime}k+q\uparrow},
\]
thus the matrix element is independent of the $k$ summation. The continuum's magnetic scattering cross section is
\begin{align*}
\frac{d\sigma}{d\Omega dE} & \propto\sum\left|\left\langle \tilde{\psi}_{l,k,\downarrow}\left|\hat{S}_{q}^{-}\right|\tilde{\psi}_{l^{\prime},k+q,\uparrow}\right\rangle \right|^{2}\delta\left(E+E_{l,k,\downarrow}-E_{l^{\prime},k+q,\uparrow}\right)\\
 & +\left|\left\langle \tilde{\psi}_{l,k,\uparrow}\left|\hat{S}_{q}^{+}\right|\tilde{\psi}_{l^{\prime},k+q,\downarrow}\right\rangle \right|^{2}\delta\left(E+E_{l,k,\uparrow}-E_{l^{\prime},k+q,\downarrow}\right)\\
 & =\sum_{ll^{\prime}k}\left|\int W_{l\downarrow}^{*}\left(r\right)W_{l^{\prime}\uparrow}\left(r\right)e^{-\mathrm{i}qr}d^{3}r\right|^{2}\delta\left(E+E_{l,k,\downarrow}-E_{l^{\prime},k+q,\uparrow}\right)\\
 & +\sum_{ll^{\prime}k}\left|\int W_{l\uparrow}^{*}\left(r\right)W_{l^{\prime}\downarrow}\left(r\right)e^{-\mathrm{i}qr}d^{3}r\right|^{2}\delta\left(E+E_{l,k,\uparrow}-E_{l^{\prime},k+q,\downarrow}\right).
\end{align*}
We can integrate over energy $E$, since the relative intensity variations versus $q$ are nearly $E$-independent. It is understood that, upon
doing so, the delta functions above (energy conservation) will be combined
with the $k$ summation under the constraint that the excitations must promote electrons from occupied to
unoccupied states, leading together to an intra-Brillouin-zone dependence of the cross section. The remaining inter-Brillouin-zone
dependence becomes:
\[
\frac{d\sigma}{d\Omega}\left(q\right)\propto\sum_{ll^{\prime}}\left|\int W_{l\downarrow}^{*}\left(r\right)W_{l^{\prime}\uparrow}\left(r\right)e^{-\mathrm{i}qr}d^{3}r\right|^{2}+\left|\int W_{l\uparrow}^{*}\left(r\right)W_{l^{\prime}\downarrow}\left(r\right)e^{-\mathrm{i}qr}d^{3}r\right|^{2}.
\]
Recall that the Wannier functions are defined as $W_{l_{d},l_{s},\sigma}\left(r-R\right)\equiv W_{l_{d},\sigma}\left(r-l_{s}-R\right)$. Again, in the limit that the MLWFs are well-localized, the leading term is obtained by summation under the constraint $l=l^{\prime}$:
\begin{align*}
\frac{d\sigma}{d\Omega}\left(q\right) & \propto\sum_{l}\left|\int W_{l}^{*}\left(r\right)W_{l}\left(r\right)e^{-\mathrm{i}qr}d^{3}r\right|^{2}\\
 & \propto\frac{1}{n_s}\sum_{l_{s}=1}^{n_s}\left|\int\left(\frac{1}{d}\sum_{l_{d}=1}^{d}\left|W_{l_{d}}\left(r-l_{s}\right)\right|^{2}\right)e^{-\mathrm{i}qr}d^{3}r\right|^{2},
\end{align*}
where $\frac{1}{d}\sum_{l_{d}=1}^{d}\left|W_{l_{d}}\left(r-l_{s}\right)\right|^{2}$
is the site-symmetric electron density of the Wannier functions. We have ignored
the spin indices in this final result because the Wannier functions for the spin-majority and minority
bands are the same if they belong to the same EBR (in the unpolarized state). This final result explains
why we can use the NNET model to account for the inter-Brillouin-zone variations of the observed neutron scattering
intensities, because: (1) the magnetic bands in MnSi form an EBR; (2) the MLWFs are centered on the entirety
of the NNETs; (3) the MLWFs are anti-bonding states, hence their electron density can be approximated by three
(``coherently superposed'') Mn ionic moments on the vertices of individual NNETs.
\newpage
\section{Supplementary Figures and Tables}
\begin{figure}[ht]
\centering{\animategraphics[width=15cm, label=Str_demo, controls]{2}{MnSi_demo4_}{01}{18}}
\caption{{\bf Animation of MnSi's crystal structure.} The animation first presents MnSi's structural derivation from rock salt. The body diagonal along which the atoms move (and its symmetry equivalent) represents the ``variable'' Wyckoff $4a$ site. Then, in the right panel, the Mn sub-lattice is viewed from a series of angles, demonstrating NNETs as a structural motif.}
\label{Figs1}
\end{figure}

\begin{figure*}[!ht]
\centering
	\includegraphics[width=1\textwidth]{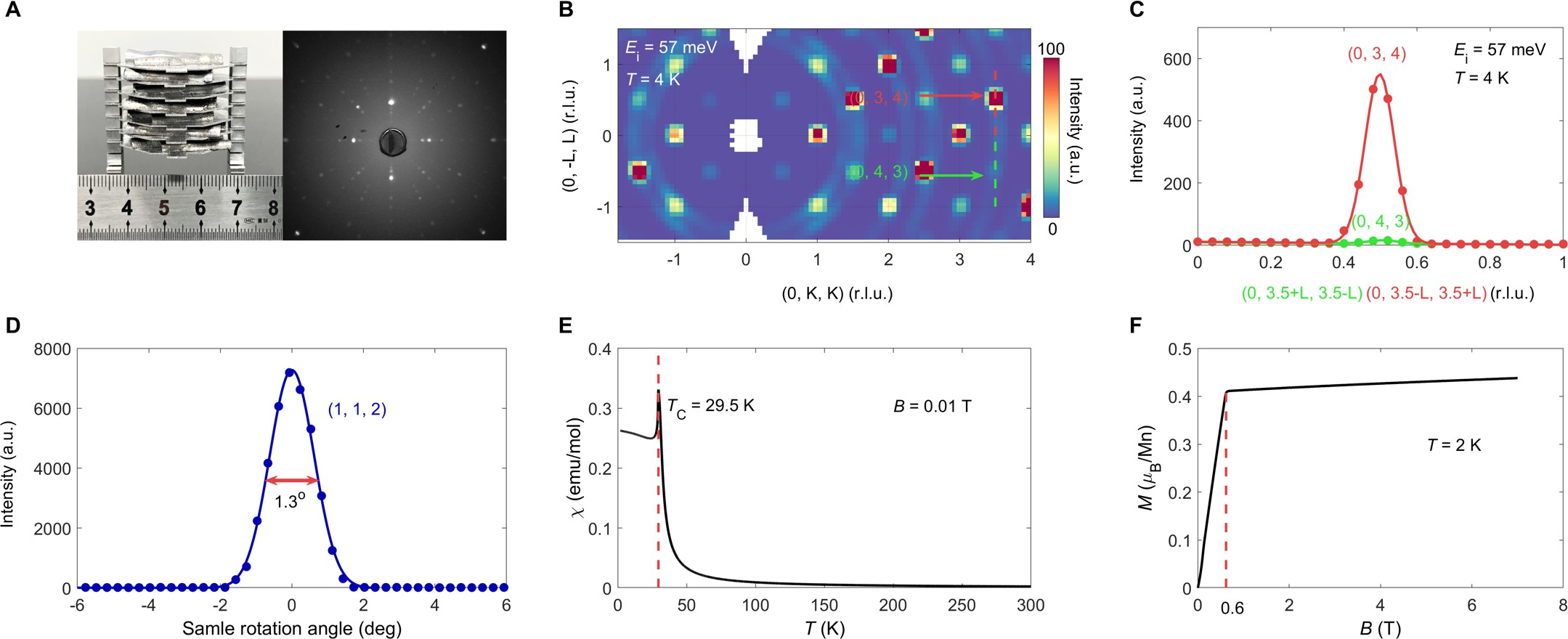}
	\caption{{\bf Sample preparation and characterization.} ({\bf A}) Left side: photograph of MnSi single crystals co-aligned on an aluminum sample holder. Right side: representative X-ray Laue pattern taken on a natural surface of a single crystal along $\langle100\rangle$ direction. ({\bf B} and {\bf C}) $\mathbf{Q}$-scan profiles which indicate that the entire sample is nearly a single-domain, since the (0, 3, 4) reflection is over 40 times more intense than that at (0, 4, 3). ({\bf D}) Neutron diffraction intensities of a selected Bragg reflection recorded upon rotating the sample. Solid line is Gaussian fit to the data, with full width at half maximum (FWHM) equal to 1.3 degrees. Data in (B)-(D) was obtained on 4SEASONS. ({\bf E} and {\bf F}) Uniform magnetization measured on a single crystal using a Quantum Design MPMS. The data indicates a helical magnetic order below 29.5 K and a saturated moment of 0.4 ${\mu}{\rm_B}$ when the magnetic field is higher than 6000 Oe, consistent with previous results \cite{WERNICK19721431}.}
	\label{Figs2}
\end{figure*}
\begin{figure*}[!ht]
\centering
	\includegraphics[width=1\textwidth]{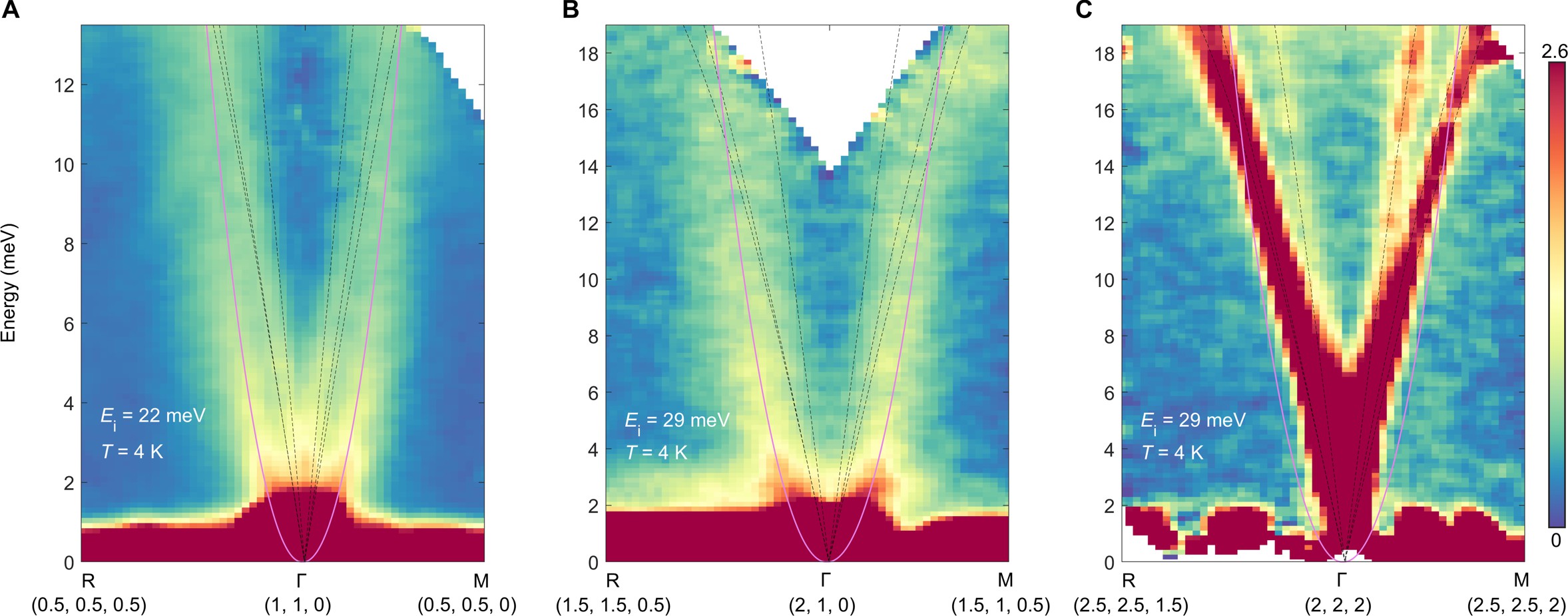}
	\caption{{\bf Distinction between spin waves and acoustic phonons.} Magenta solid lines indicate the spin waves, whereas black dashed lines indicate the acoustic phonon dispersions. The spin waves are most clearly seen in the (2, 1, 0) BZ. Data were obtained on 4SEASONS.}
	\label{Figs3}
\end{figure*}
\begin{figure*}[!ht]
\centering
	\includegraphics[width=0.5\textwidth]{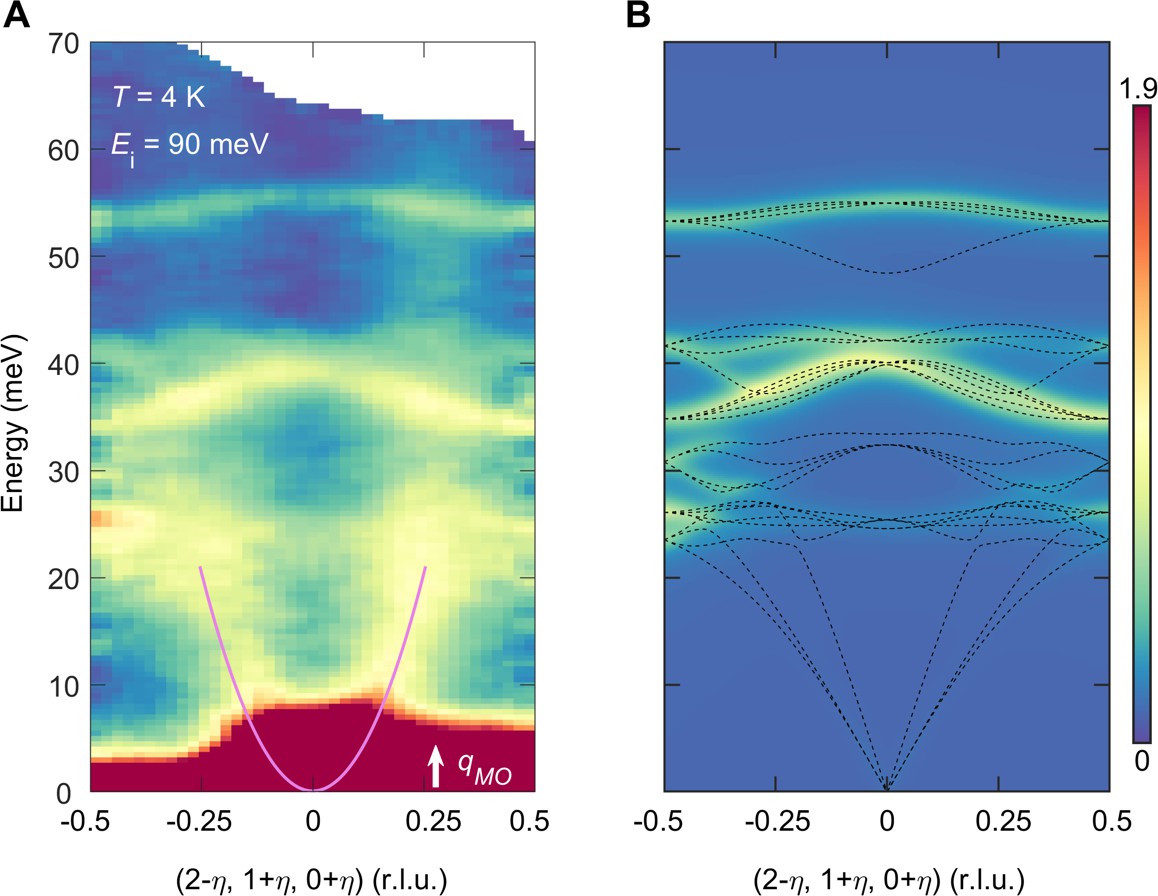}
	\caption{{\bf Comparison between INS and DFT-calculated phonon spectra in the (2, 1, 0) BZ.} Data were obtained on 4SEASONS.}
	\label{Figs4}
\end{figure*}
\begin{figure*}[!ht]
\centering
	\includegraphics[width=1\textwidth]{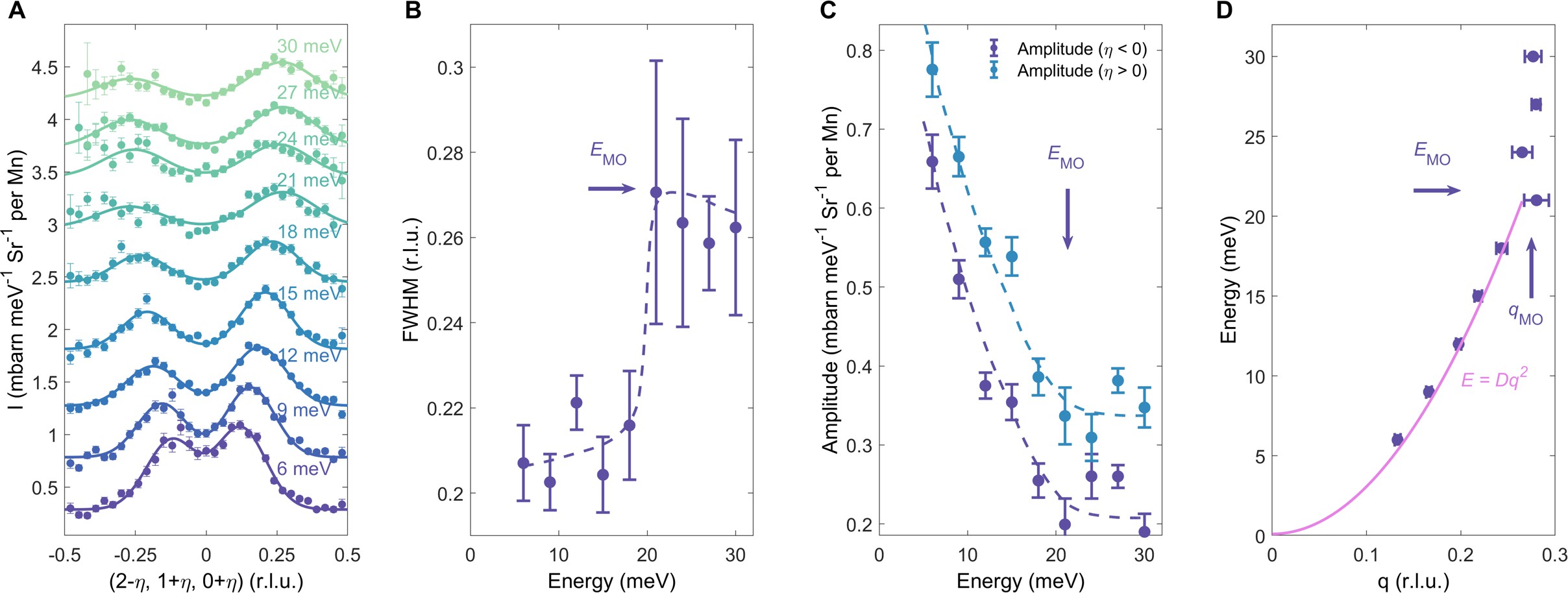}
	\caption{{\bf Momentum (and energy) evolution of the spin excitations.} ({\bf A}) Constant-$E$ cuts from 6 meV to 30 meV with $dE = \pm$ 0.7 meV collected at $T$ = 4 K and ${E}_{\rm i}$ = 57 meV on 4SEASONS, fitted with two symmetrically positioned Gaussian peaks on a constant background. ({\bf B}-{\bf D}) Fit parameters displayed as a function of the cut energy in (A). Error bars represent one standard deviation of the data. The spectrum undergoes an abrupt change at $q_\mathrm{MO}$, which corresponds to an energy of $E_\mathrm{MO}$ of about 20 meV, where the spin waves meet the nearly $E$-independent continuum.}
	\label{Figs5}
\end{figure*}
\begin{figure*}[!ht]
\centering
	\includegraphics[width=1\textwidth]{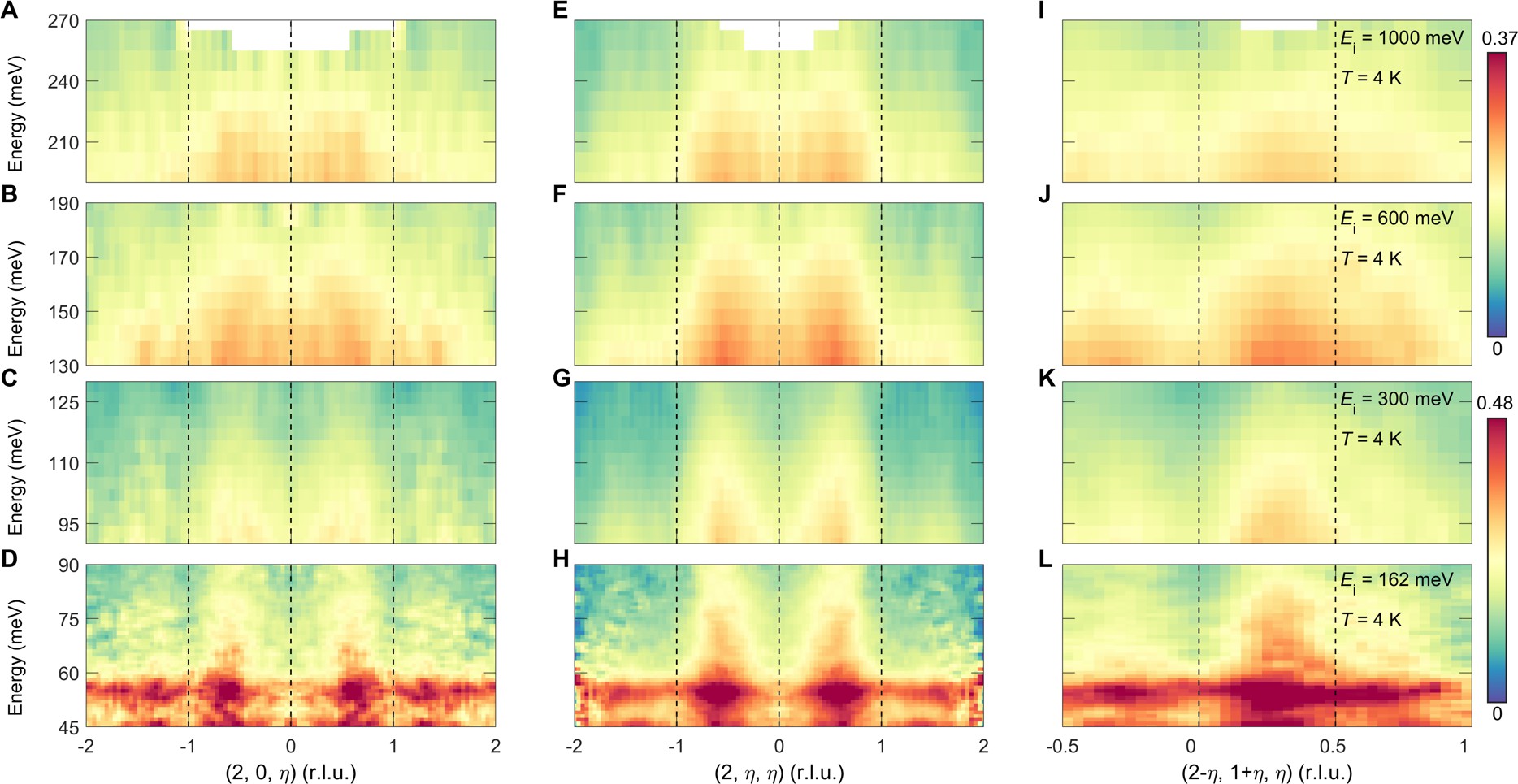}
	\caption{{\bf Representative $E$-$\mathbf{Q}$ slices along three high-symmetry directions.} ({\bf A}-{\bf L}) The continuum's characteristic $\mathbf{Q}$-dependence is nearly independent of energy. Data were obtained at $T$ = 4 K and with ${E}_{\rm i}$ = 162 (4SEASONS), 300, 600 and 1000 meV (ARCS).}
	\label{Figs6}
\end{figure*}
\begin{figure*}[!ht]
\centering
	\includegraphics[width=1\textwidth]{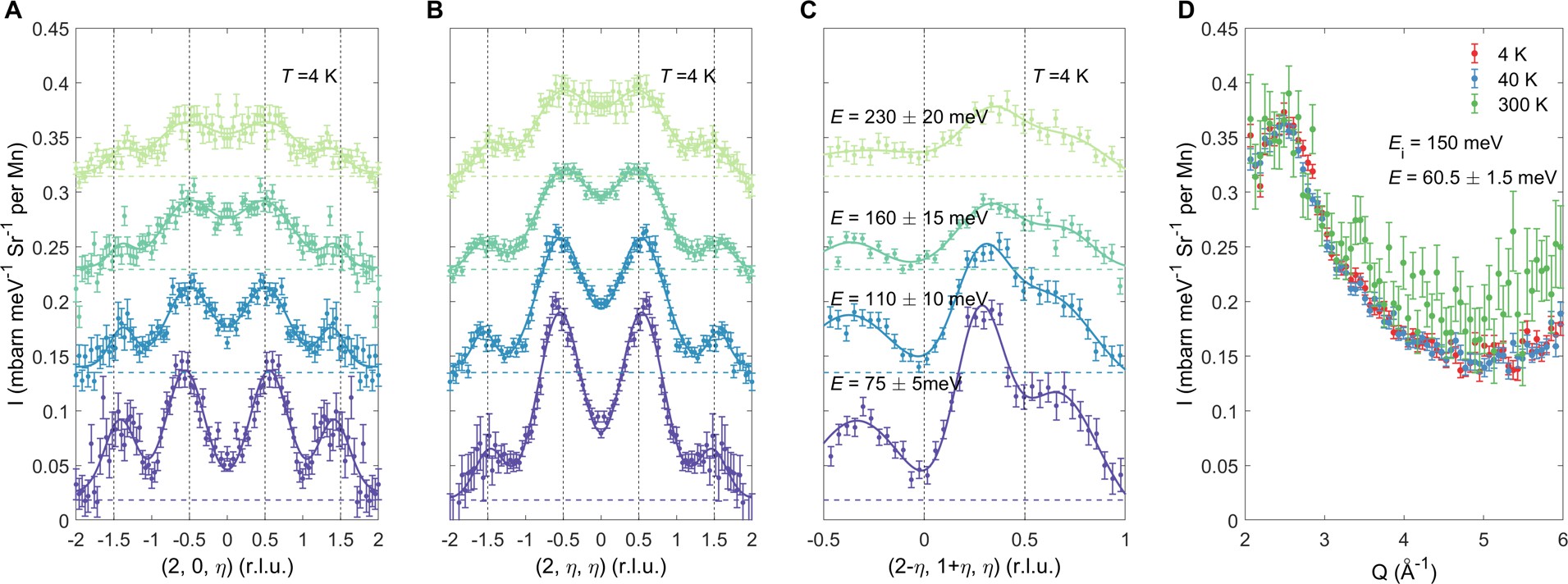}
	\caption{{\bf Constant-$E$ cuts along three high-symmetry directions and comparison of INS intensity of the continuum between different temperatures.} ({\bf A}-{\bf C}) Data are fit with Gaussian peaks with a constant background, offset for clarity. The measurements were performed at $T$ = 4 K and with ${E}_{\rm i}$ = 162 (4SEASONS), 300, 600 and 1000 meV (ARCS). ({\bf D}) Radial $Q$-dependent intensities measured with ${E}_{\rm i}$ = 150 meV (SEQUOIA). Error bars represent one standard deviation of the counting statistics.}
	\label{Figs7}
\end{figure*}
\begin{figure*}[!ht]
\centering
	\includegraphics[width=1\textwidth]{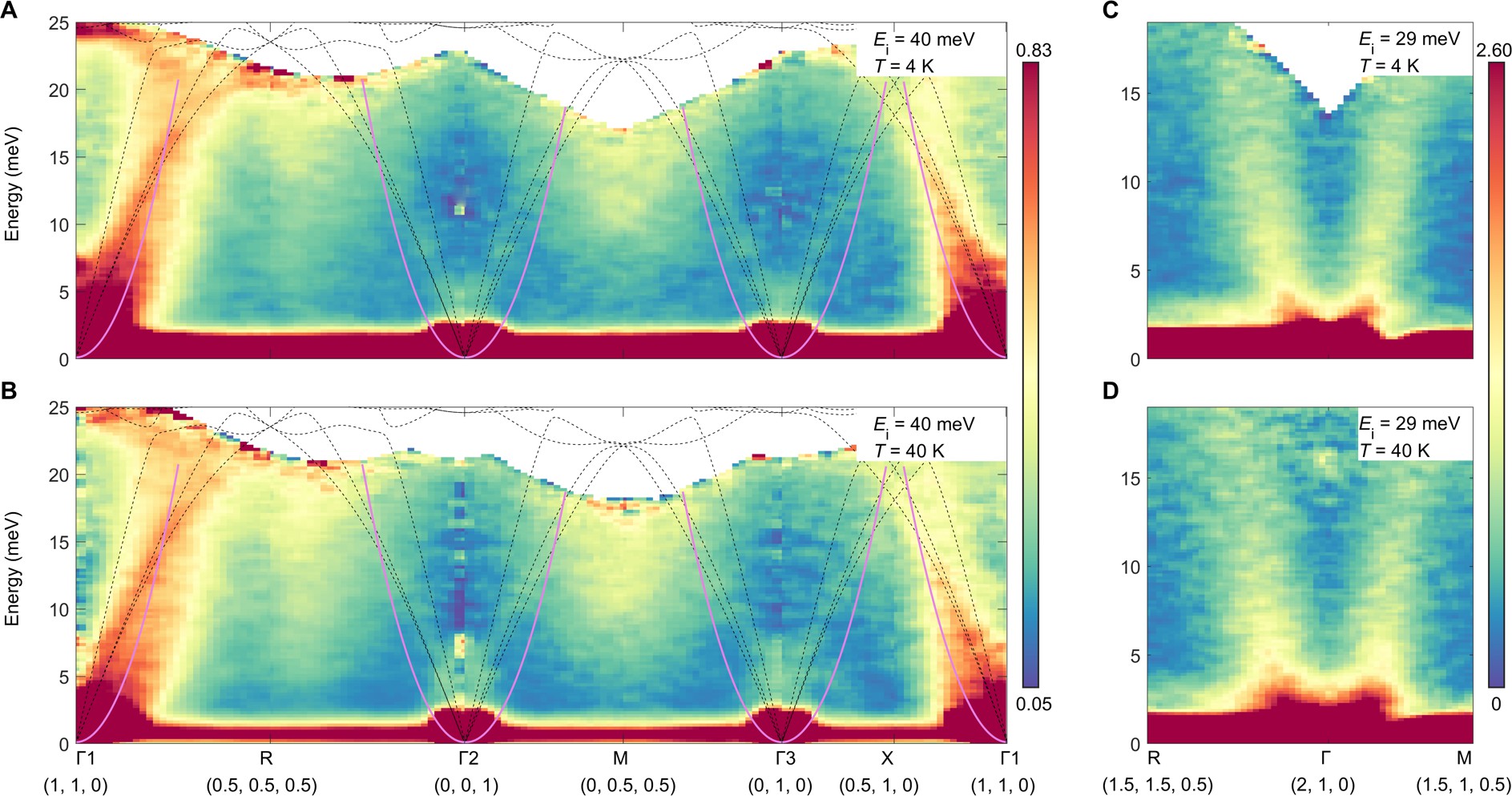}
	\caption{{\bf INS signal variation between 4 K and 40 K.} The detailed balance factor has been divided from the data. It is seen that both the continuum and the spin waves (paramagnons) persist to above $T_\mathrm{N}$. Measurements were performed on 4SEASONS.}
	\label{Figs8}
\end{figure*}
\begin{figure*}[!ht]
\centering
	\includegraphics[width=0.5\textwidth]{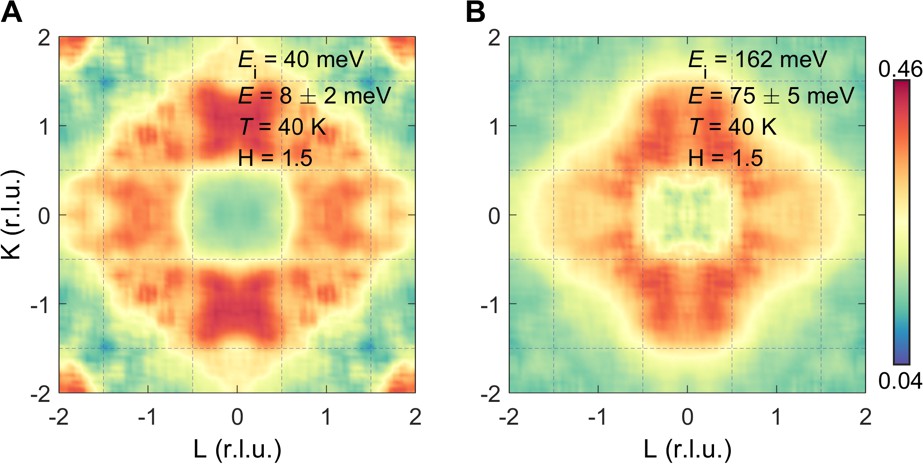}
	\caption{{\bf Comparison of intensity variations over the same set of Brillouin zones for the low- and high-energy parts of the continuum.} Thin dashed lines indicate Brillouin-zone boundaries. Data were obtained on 4SEASONS.}
	\label{Figs9}
\end{figure*}

\begin{figure*}[!ht]
\centering
	\includegraphics[width=0.5\textwidth]{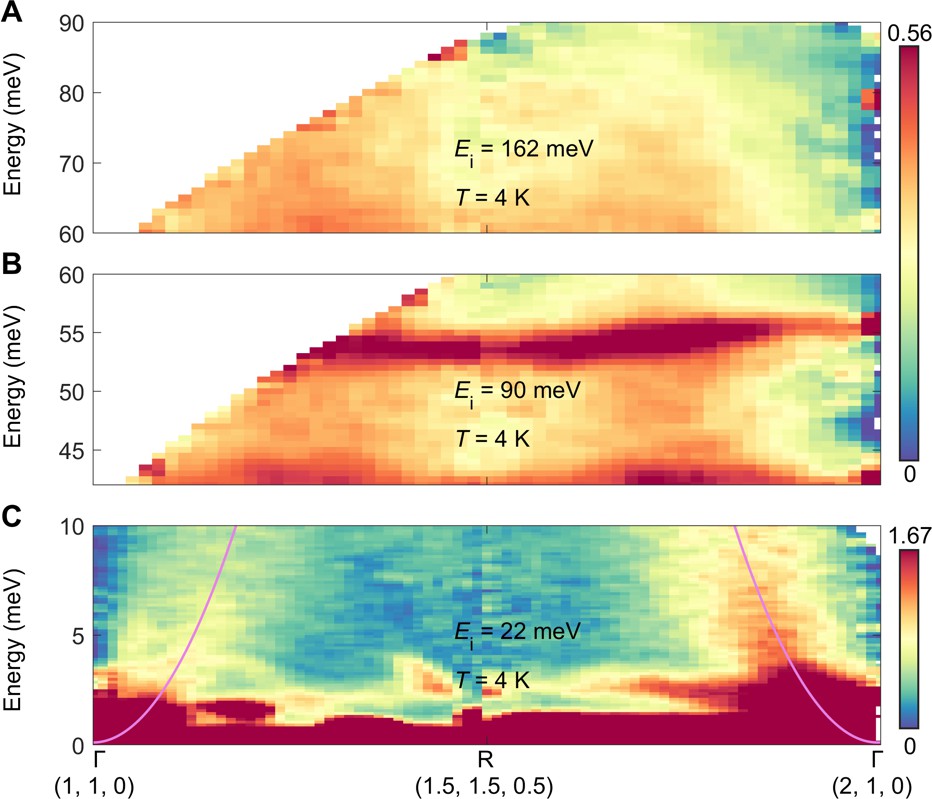}
	\caption{{\bf Relative variation of INS intensity of the spin waves and high-energy continuum in neighboring BZs.} The intensity of the spin waves in the (1, 1, 0) BZ is much weaker than that in the (2, 1, 0) BZ, but the INS intensity of continuum above 45 meV in the (1, 1, 0) BZ is stronger than that in the (2, 1, 0) BZ, indicating their different microscopic origin. Data were obtained on 4SEASONS.}
	\label{Figs10}
\end{figure*}

\begin{figure*}[!ht]
\centering
	\includegraphics[width=1\textwidth]{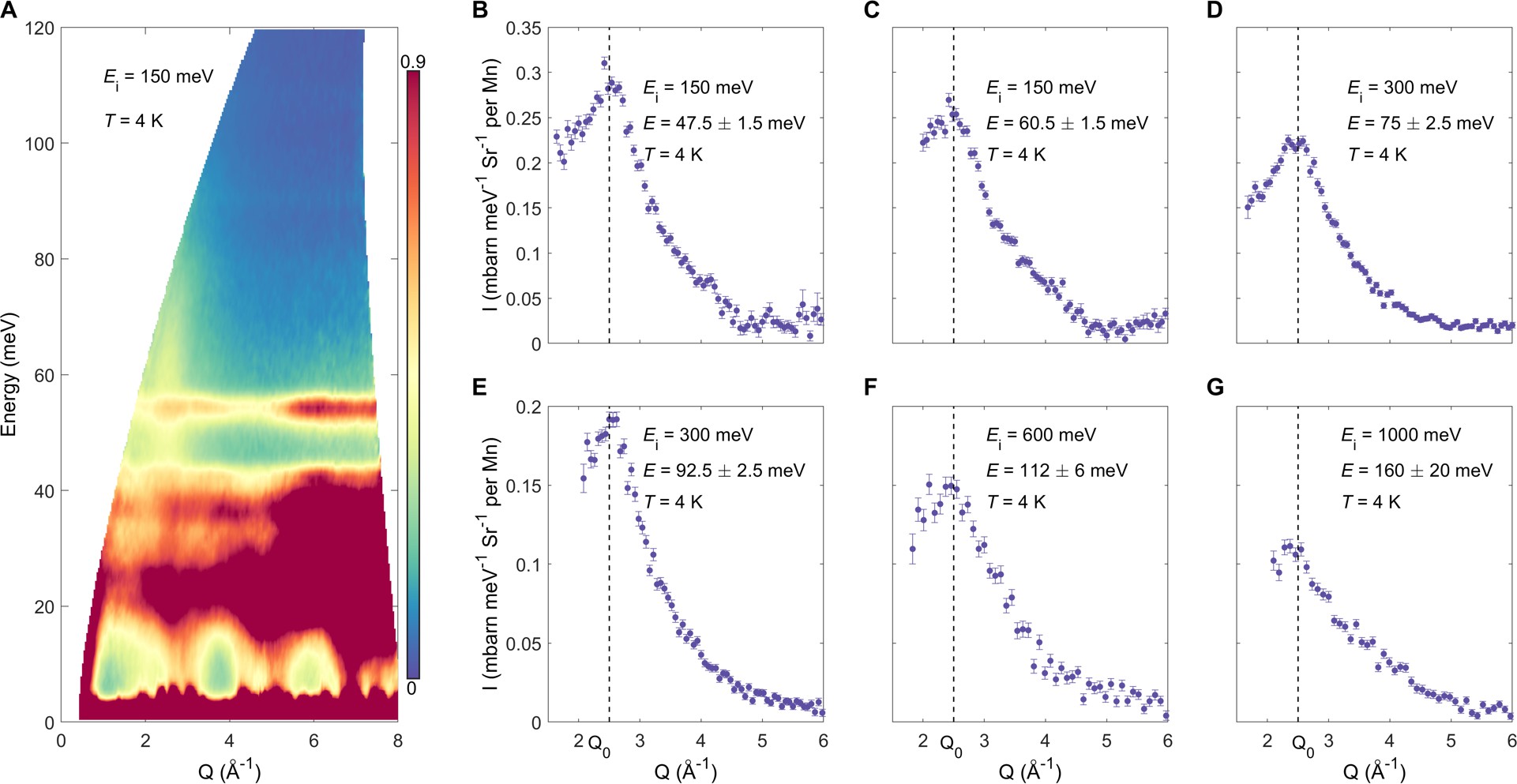}
	\caption{{\bf Radial momentum dependence of INS intensity of the continuum.} ({\bf A}) $E - Q$ dependence of the INS intensity. ({\bf B}-{\bf G}) Radial $Q$ dependence of signal extracted from constant-$E$ radial-$Q$ cuts that have avoided phonon scattering. The local maximum at $Q_{\rm 0} = 2.5~{\rm\AA^{-1}}$ is observed in all the data. The fitted non-magnetic background has been subtracted (see Eq.~\ref{EqS2}). Data were collected at $T = \rm 4~K$ with $E_{\rm i} = \rm 150~meV$ (SEQUOIA), 300 meV, 600 meV and 1000 meV (ARCS).}
	\label{Figs11}
\end{figure*}
\begin{figure*}[!ht]
\centering
	\includegraphics[width=1\textwidth]{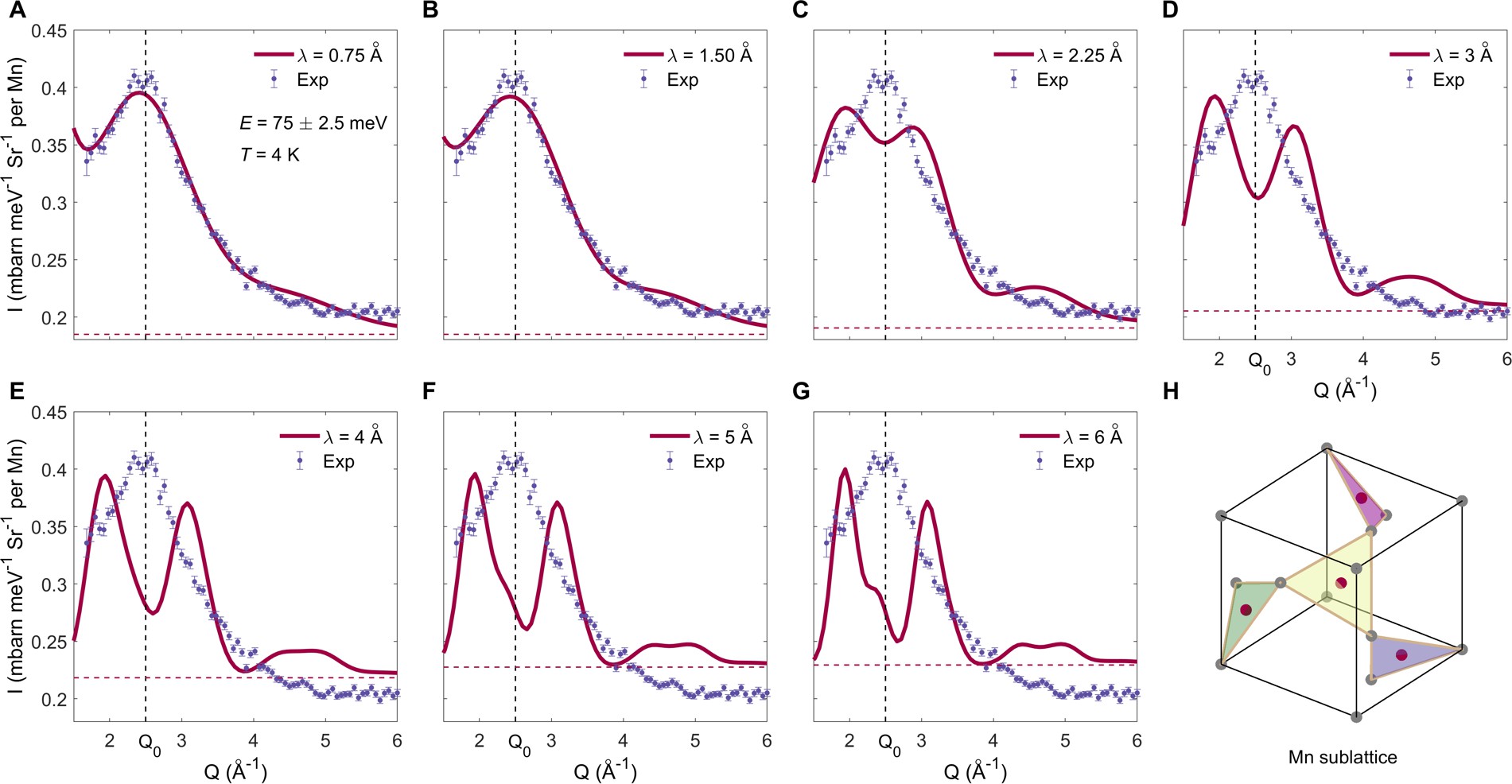}
	\caption{{\bf Simulation results with correlation center (see Eq.~\ref{EqS1}) located at the centers of NNETs under different correlation length.} ({\bf A}-{\bf G}) Best fit results using the least squares method with constant background. ({\bf H}) The Mn sublattice with 4 types of NNETs, on which the centers of NNETs are denoted with red dots. When the correlation length exceeds the size of a triangle, the modeled cluster becomes more extensive than a single NNET and the agreement with the experimental data becomes much poorer. Data were collected with $E_{\rm i} = \rm 300~meV$ (ARCS). Error bars represent one standard deviation of the counting statistics.}
	\label{Figs12}
\end{figure*}

\begin{figure*}[!ht]
\centering
	\includegraphics[width=1\textwidth]{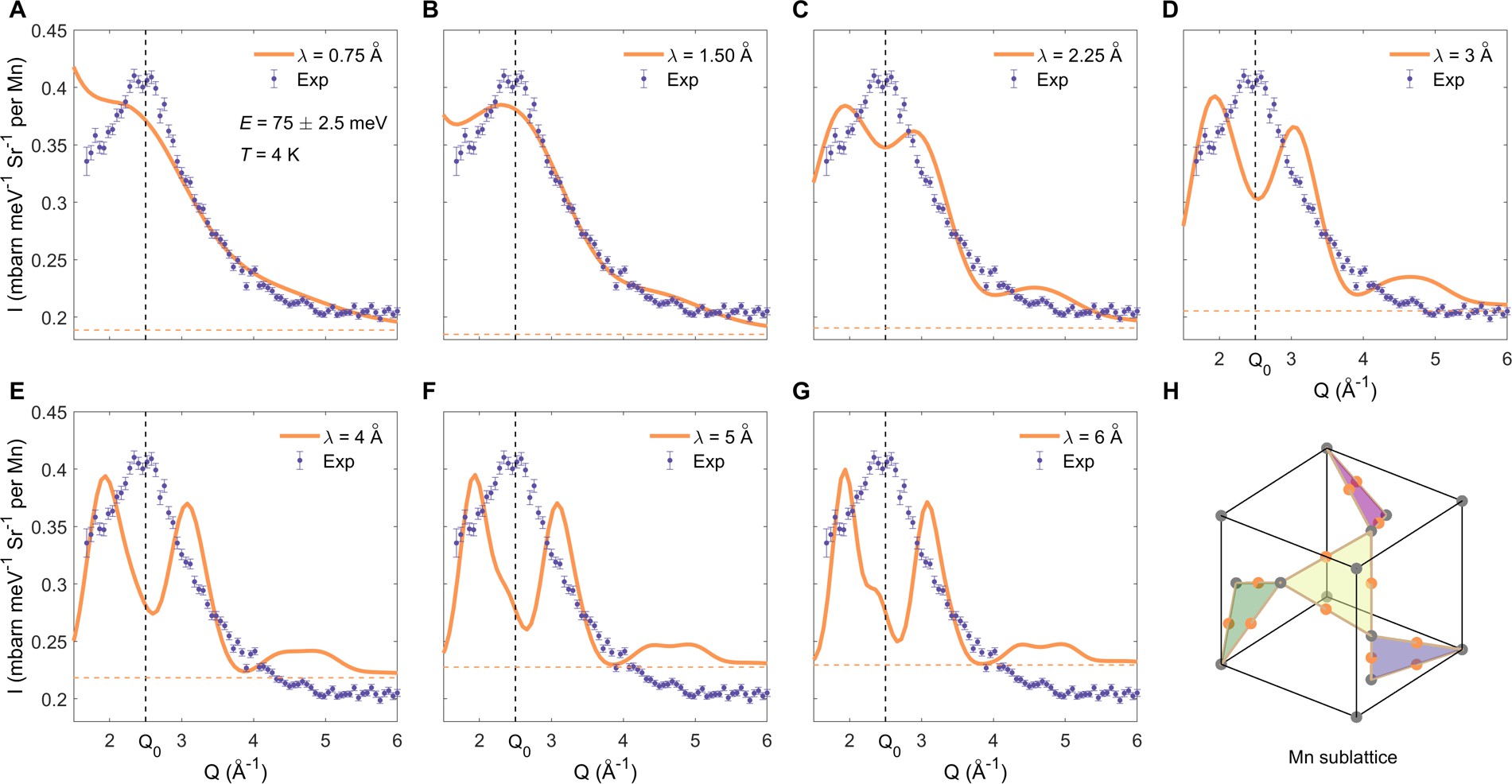}
	\caption{{\bf Simulation results with correlation center (see Eq.~\ref{EqS1}) located at the midpoint of two nearest Mn ions under different correlation length.} ({\bf A}-{\bf G}) Best fit results using the least squares method with constant background. ({\bf H}) The Mn sublattice with 4 types of NNETs, on which the midpoints of nearest Mn ions are denoted with orange dots. When the correlation length is smaller than half of the nearest Mn--Mn distance (1.4 $\mathrm{\AA}$), the cluster is essentially an Mn--Mn diatomic group. Data were collected with $E_{\rm i} = \rm 300 meV$ (ARCS). Error bars represent one standard deviation of the counting statistics.}
	\label{Figs13}
\end{figure*}
\begin{figure*}[!ht]
\centering
	\includegraphics[width=1\textwidth]{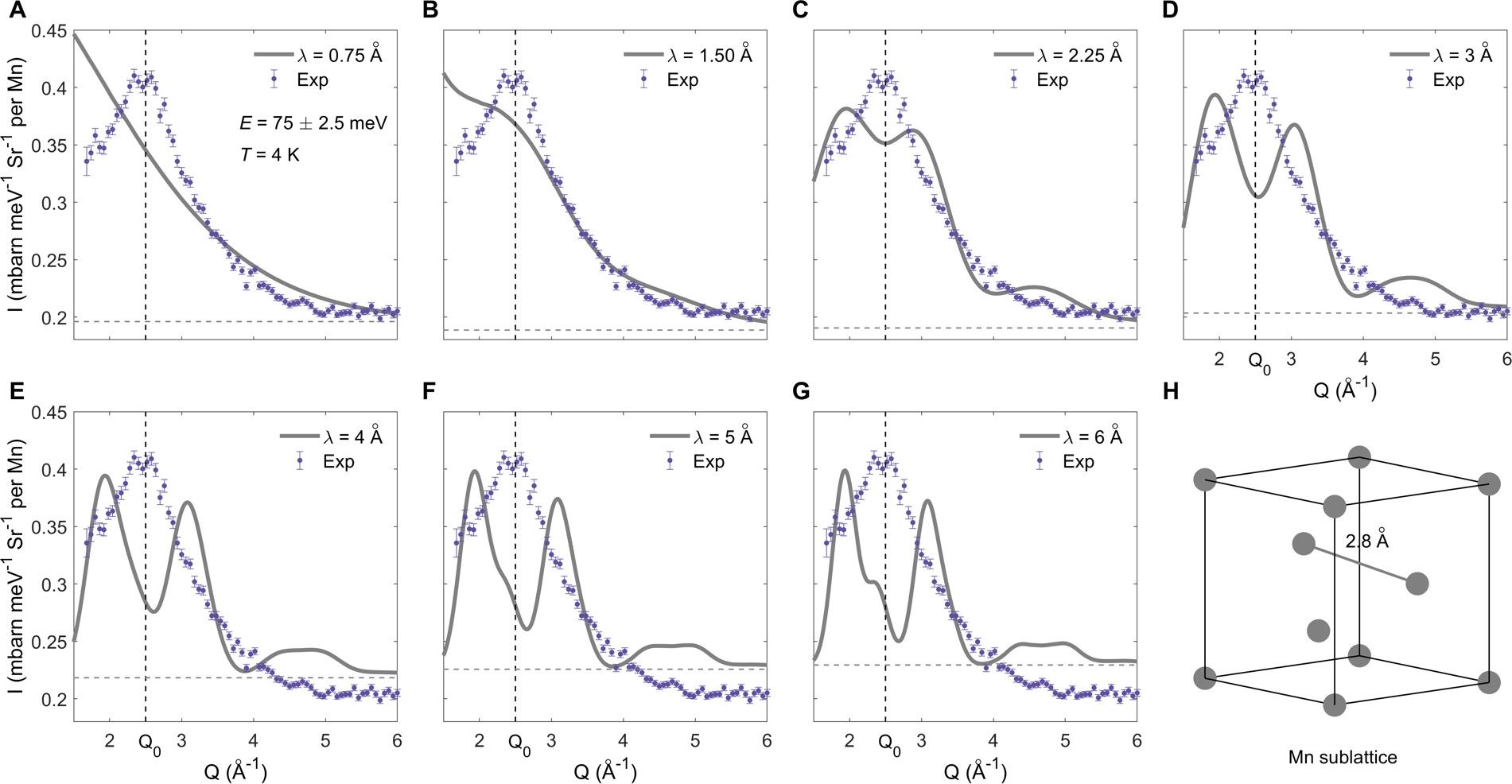}
	\caption{{\bf Simulation results with correlation center (see Eq.~\ref{EqS1}) located at the Mn ions under different correlation length.} ({\bf A}-{\bf G}) Best fit results using the least squares method with constant background. ({\bf H}) The Mn sublattice. The distance of nearest Mn--Mn atoms is 2.8 $\mathrm{\AA}$. Data were collected with $E_{\rm i} = \rm 300 meV$ (ARCS). Error bars represent one standard deviation of the counting statistics.}
	\label{Figs14}
\end{figure*}

\begin{figure*}[!ht]
\centering
	\includegraphics[width=1\textwidth]{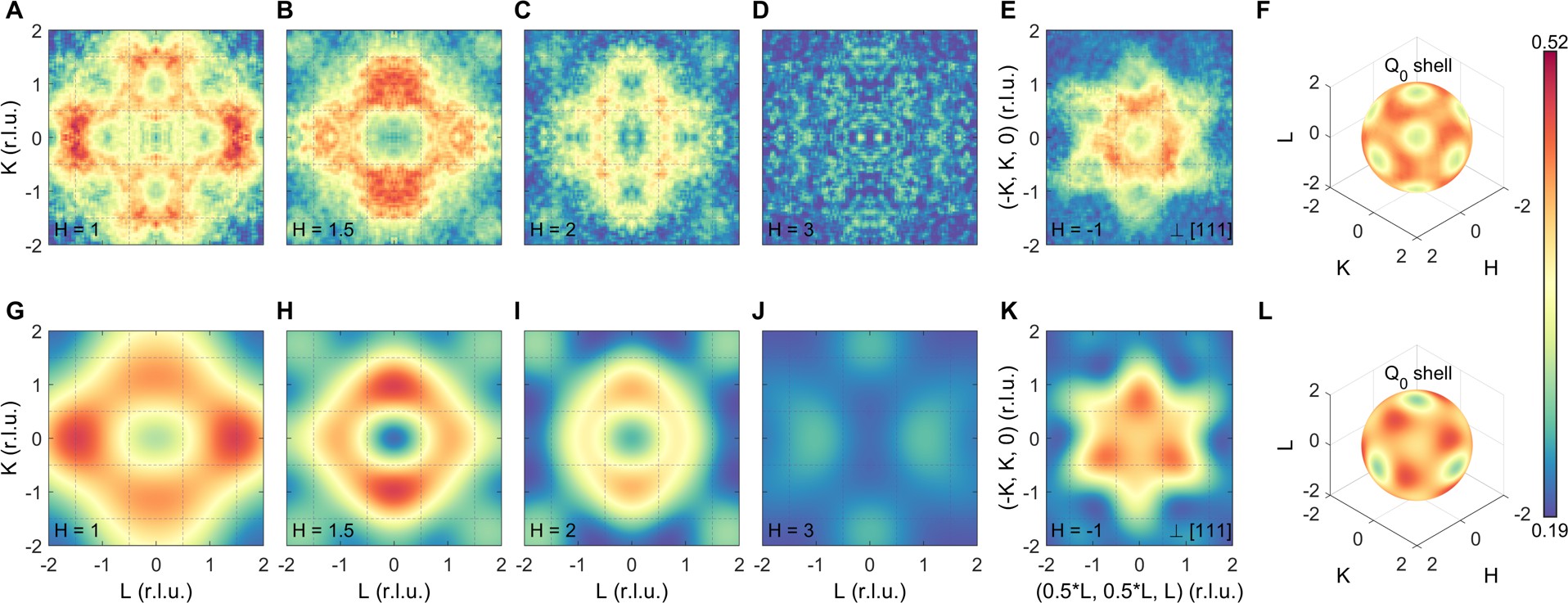}
	\caption{{\bf Comparison between experimental data and simulation results described in Eq.~\ref{EqS2}} ({\bf A}-{\bf D}) INS intensity of the $KL$ plane for $H$ = 1, 1.5, 2, 3, respectively. ({\bf E}) INS intensity at a plane which is perpendicular to [111] direction. ({\bf F}) INS intensities on the ${Q}_{\rm 0}$ shell. ({\bf G}-{\bf L}) The simulation results under the same conditions using Eq.~\ref{EqS2}, \textit{i.e.}, without considering intra-BZ models (shell or lattice-Lorentzian). All the experimental data were obtained at $T$ = 4 K, $E = 75~\pm~10$ meV and with ${E}_{\rm i}$ = 300 meV (ARCS).}
	\label{Figs15}
\end{figure*}

\begin{figure*}[!ht]
\centering
	\includegraphics[width=1\textwidth]{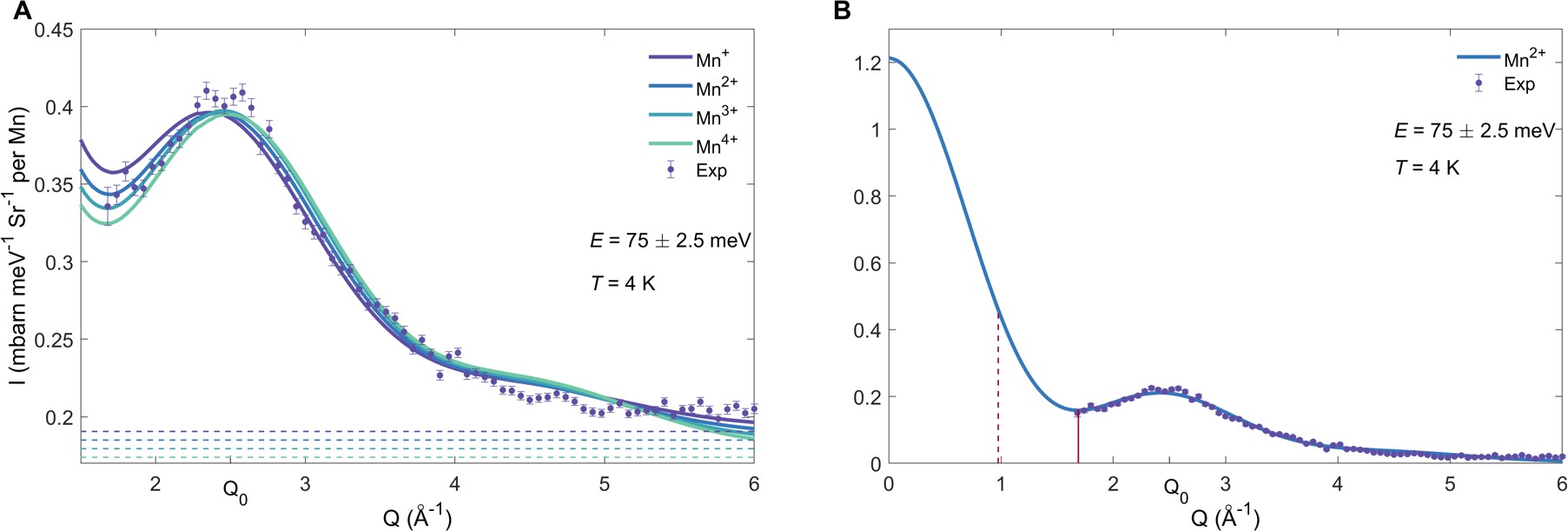}
	\caption{{\bf Comparison between the experimental data and calculation results using Eq.~\ref{EqS2} (or \ref{EqS3}, \ref{EqS5}, nearly no difference) with ionic form factor $f(Q)$ for different valence states of Mn.} ({\bf A}) The best agreement with INS data is obtained with Mn$^{2+}$. ({\bf B}) Same as (A), but after subtracting the fit background and extrapolated to $Q=0$. The red dashed and solid lines indicate the $Q$ lengths of $\mathrm{M_1}$ = (0, 0.5, 0.5) and $\mathrm{M_2}$ = (1, 0.5, 0.5) in Fig.~\ref{Figs17}, respectively.  Error bars represent one standard deviation of the counting statistics.}
	\label{Figs16}
\end{figure*}

\begin{figure*}[!ht]
\centering
	\includegraphics[width=0.6\textwidth]{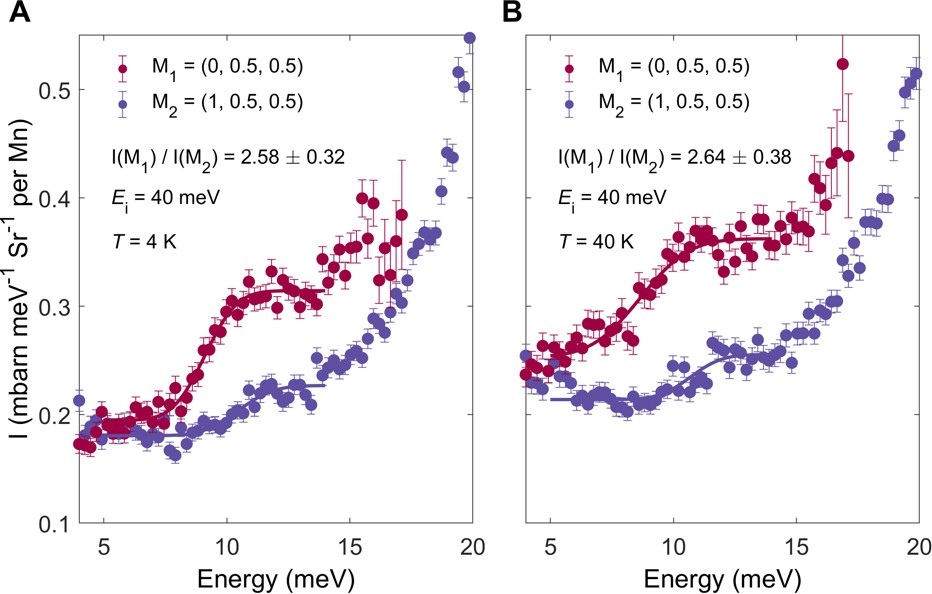}
	\caption{{\bf Relative variation of intensity of the continuum in its low-energy limit, measured at two different M points.} The data (obtained on 4SEASONS) are fitted with a step-like function on a constant background, in order to extract the intensity of the continuum at its energy minimum. \it I\rm(M) corresponds to the height of the step function. Error bars represent one standard deviation of the counting statistics. The calculated NNET form factor ratio between M$_1$ and M$_2$ is 2.56, in agreement with the measurements.}
	\label{Figs17}
\end{figure*}

\begin{figure*}[!ht]
\centering
	\includegraphics[width=1\textwidth]{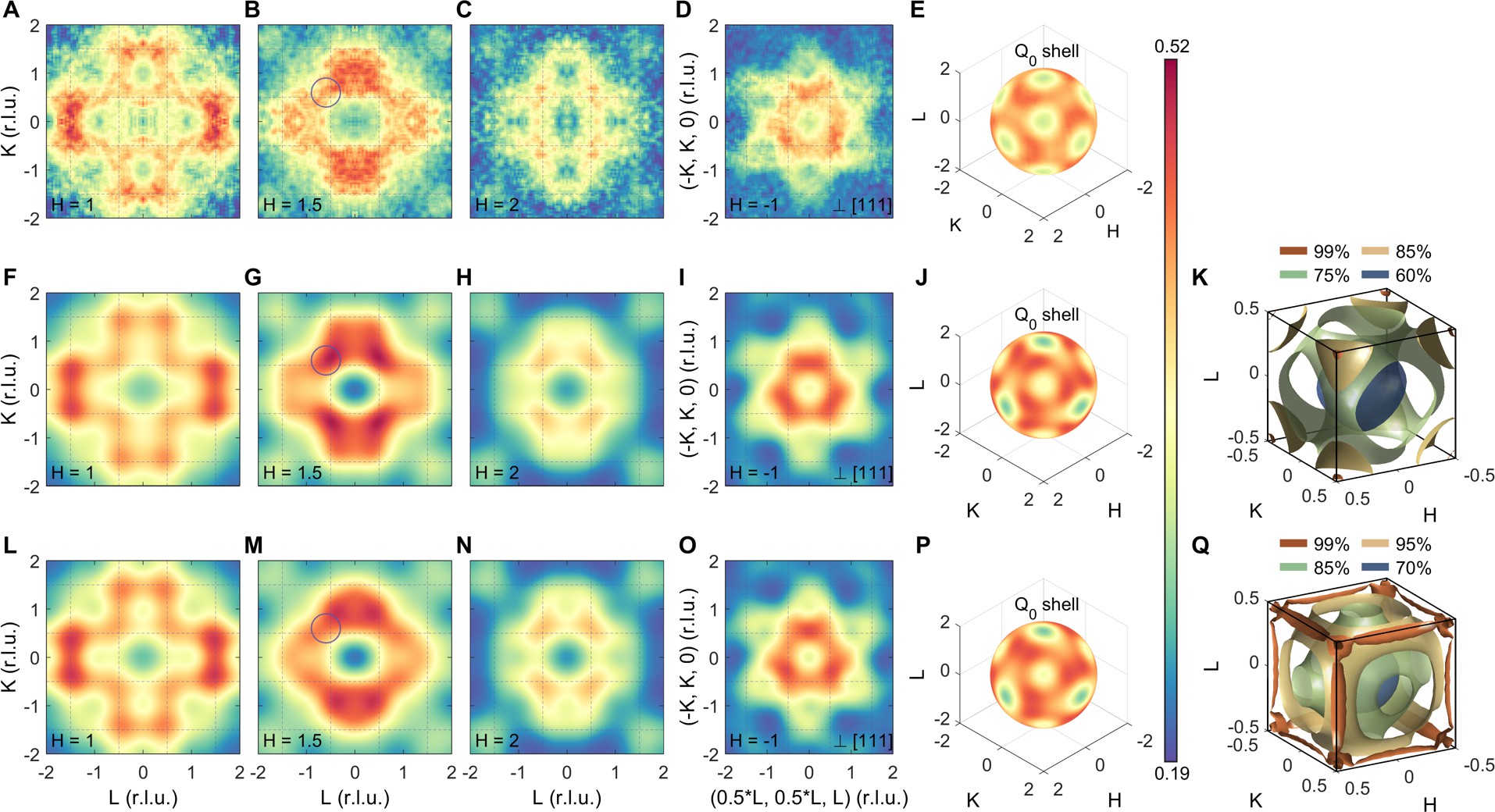}
	\caption{{\bf Comparison between experimental data and simulation results described in Eq.~\ref{EqS3} and \ref{EqS5}.} ({\bf A}-{\bf C}) INS intensity of $KL$ plane for $H$ = 1, 1.5, 2, respectively. ({\bf D}) INS intensity at a plane which is perpendicular to [111] direction. ({\bf E}) INS intensities on the ${Q}_{\rm 0}$ shell. ({\bf F}-{\bf J}) Simulation results described in Eq.~\ref{EqS3}. ({\bf K}) Isosurfaces of intra-BZ intensity described by the normalized lattice-Lorentzian function ($\rm{LL}(\bf Q)$). ({\bf L}-{\bf P}) Simulation results described in Eq.~\ref{EqS5}. ({\bf Q}) Isosurfaces of intra-BZ intensity described by the spherical shell function ($\rm{Shell}(\bf Q)$). Experimental data were obtained at $T$ = 4 K, $E = 75 \pm 10$ meV and with ${E}_{\rm i}$ = 300 meV (ARCS). Circles in (B), (G) and (M) indicate where the shell model describes the experiment slightly better than the lattice-Lorentzian model, but apart from that the two models are highly comparable.}
	\label{Figs18}
\end{figure*}

\begin{figure*}[!ht]
\centering
	\includegraphics[width=0.6\textwidth]{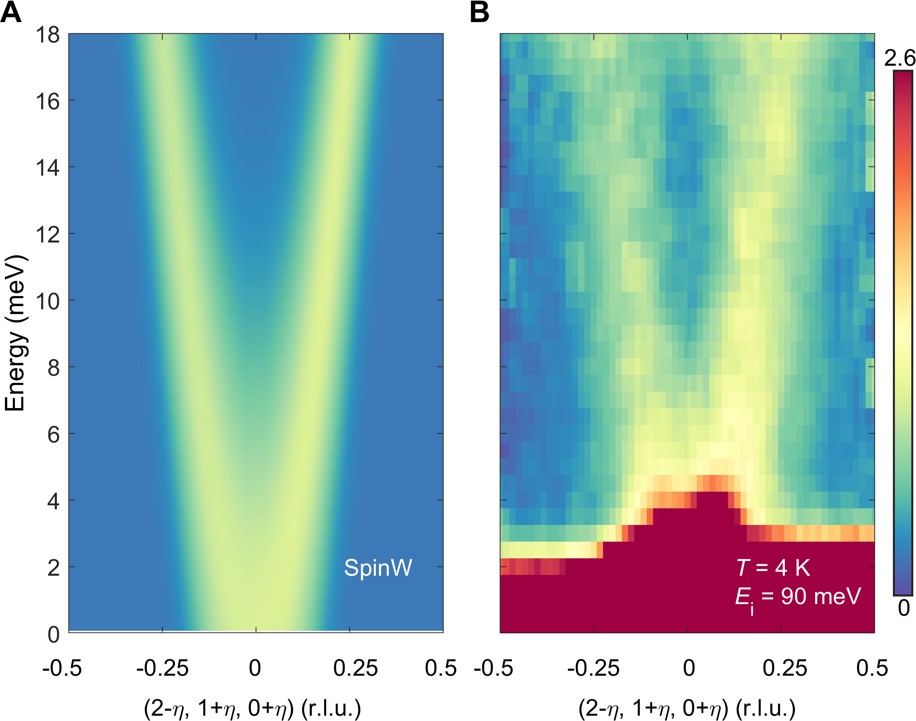}
	\caption{{\bf Comparison between INS and SpinW-calculated acoustic spin-wave spectra in the (2, 1, 0) BZ.} Experimental data were obtained on SEQUOIA.}
	\label{Figs19}
\end{figure*}

\begin{figure*}[!ht]
\centering
	\includegraphics[width=0.6\textwidth]{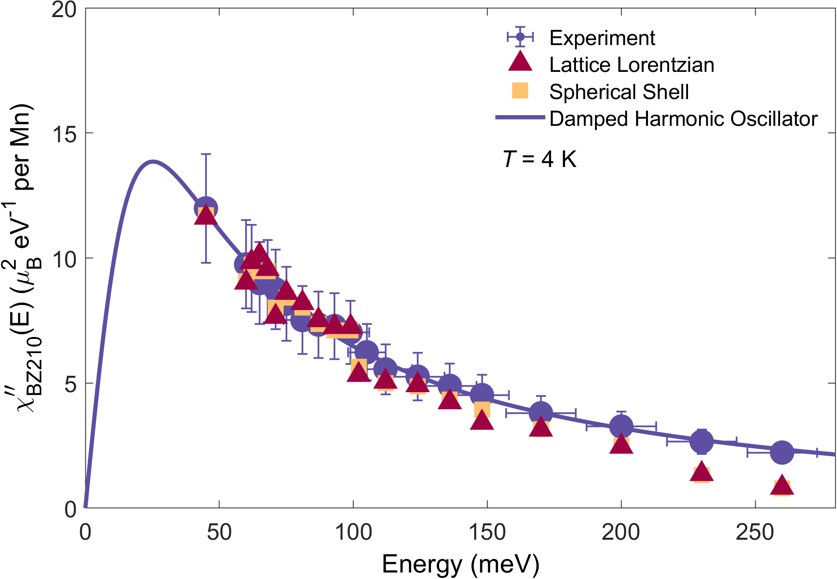}
	\caption{{\bf Dynamic magnetic susceptibility determined from experiment as a function of energy.} Solid line is fit to the $\chi^{\prime\prime}(E)$ of a damped harmonic oscillator. Squares and triangles are fit intensities based on spherical shell model and lattice Lorentzian model, and all data are integrated over the (2,~1,~0) BZ. Data were obtained at ${E}_{\rm i}$ = 150 meV (SEQUOIA), 300 meV, 600 meV and 1000 meV (ARCS). Error bar represent the joint uncertainty associated with the counting statistics, vanadium calibration and neutron-absorption calculation.}
	\label{Figs20}
\end{figure*}

\begin{figure*}[!ht]
\centering
	\includegraphics[width=0.7\textwidth]{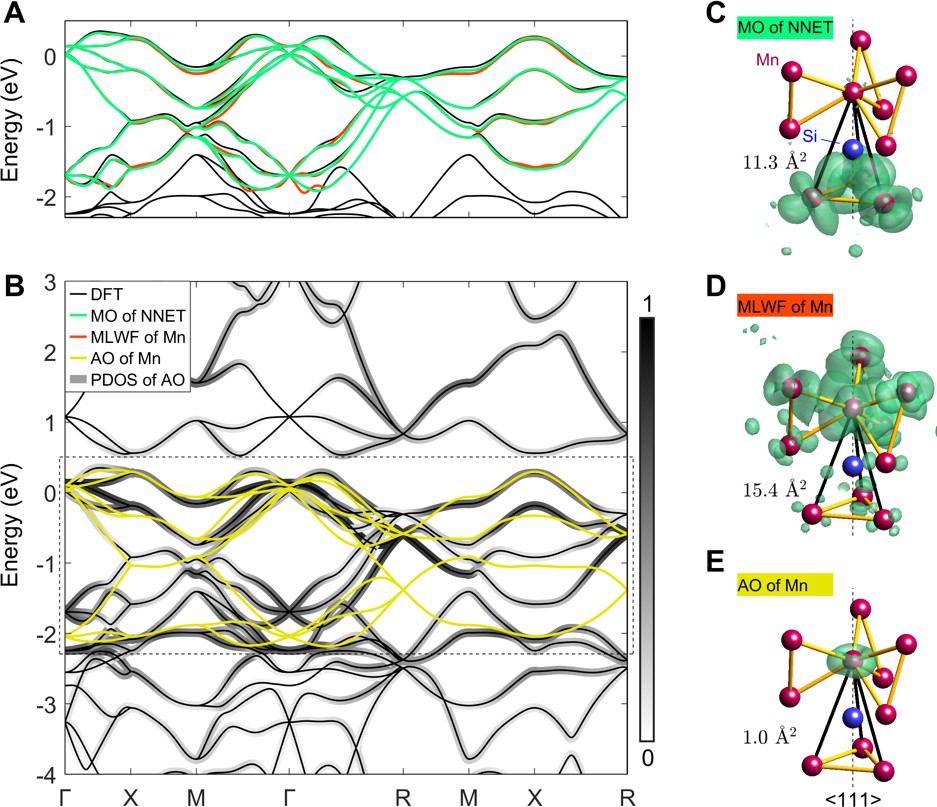}
	\caption{\textbf{\textit{Ab initio} calculation of magnetic orbitals.} (\textbf{A}) and (\textbf{B}) DFT-calculated bands near $E_\mathrm{F}$ in comparison to TB models. All legends are displayed in (B). For clarity, only spin-majority bands are shown (minority bands in Fig.~\ref{Figs23}). Three types of orbitals are compared for their effectiveness to describe the magnetic EBR at $E_\mathrm{F}$ (see text): (1) MLWFs calculated from the EBR without site constraint, resulting in MOs on NNETs. (2) MLWFs calculated from the EBR under Mn-site constraint. (3) MLWFs calculated from a total of 32 bands near $E_\mathrm{F}$, resulting in approximate AOs, among which the displayed $e_{x^2-y^2}$ and $e_{xy}$ orbitals of Mn ($z\equiv$ the local $C_3$ axis) provide the closest description of the EBR, but deviation is significant and the PDOS enters bands outside the EBR. (\textbf{C}-\textbf{E}) Isosurface of the corresponding electron density that encloses 50\% of the electron weight, shown on a structural motif. The spreads of electron density $\left\langle r^2 \right\rangle - \left\langle r \right\rangle^2$ \cite{MarzariRMP2012} are 11.3, 15.4, and 1.0~$\mathrm{\AA}^2$, respectively.}
	\label{Figs21}
\end{figure*}

\begin{figure*}[!ht]
\centering
	\includegraphics[width=1\textwidth]{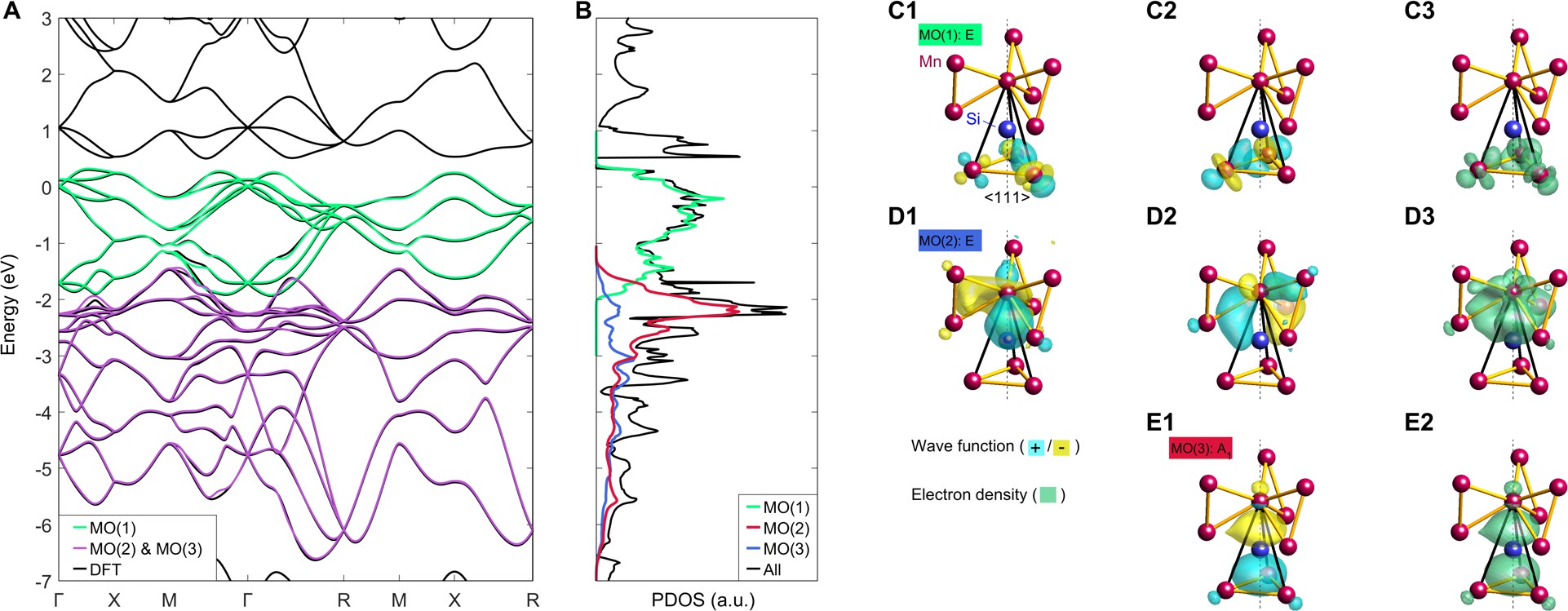}
	\caption{{\bf A molecular-orbital description of occupied electronic states.} (\textbf{A}) DFT-calculated spin-majority bands near $E_\mathrm{F}$, along with TB-model interpolated bands according to MLWF calculations. Two sets of bands ($8 + 12$) are seen within $\sim6$ eV below $E_\mathrm{F}$. The 8 bands closer to $E_\mathrm{F}$ are the magnetic EBR discussed in the main text. The remaining 12 bands form a BR (not an EBR) which is jointly induced by an $E$ and an $A_1$ irrep of the Wyckoff $4a$ site. MLWF calculation using this BR's Bloch states as input results in a satisfactory TB model, the local orbitals of which has the corresponding symmetries. These orbitals are MOs as well, because their spatial extent is much greater than atomic orbitals. All of the EBR- (and BR-) restricted MLWFs (total of 20), MO(1), MO(2) and MO(3), are mutually orthogonal, and they are centerd on the NNET, Mn, and Si sites (all Wyckoff $4a$), respectively. These MOs form because of the covalent bonding nature between Mn and Si, together with the fact that every Mn is surrounded by a total of seven Si (Fig.~\ref{Figs1}) at similar distances (ranging from 2.3 to 2.5 $\mathrm{\AA}$) in the crystal structure. (\textbf{B}) Partial density of states of the MOs. It is seen that while MO(1) is exclusively responsible for the 8-band EBR near $E_\mathrm{F}$, MO(2) and MO(3) jointly form the 12-band BR, as expected. (\textbf{C1}-\textbf{C3}) Wave functions and symmetrised electron density isosurfaces of MO(1) (isovalues correspond to $15\%$ of the maximal density), displayed in the ``trillium flower'' structural motif. (\textbf{D1}-\textbf{D3}) Same as (C1)-(C3), but for MO(2). (\textbf{E1}-\textbf{E3}) Same as (C1)-(C3), but for MO(3), which is a one-dimensional irrep (hence, only one wave function). With the understanding that all Mn atoms and NNETs are symmetry-related, MO(1), MO(2) and MO(3) can be named after the shapes of their electron clouds as (trillium) ``root'', ``flower'' and ``stem'' MOs, respectively.}
	\label{Figs22}
\end{figure*}

\begin{figure*}[!ht]
\centering
	\includegraphics[width=1\textwidth]{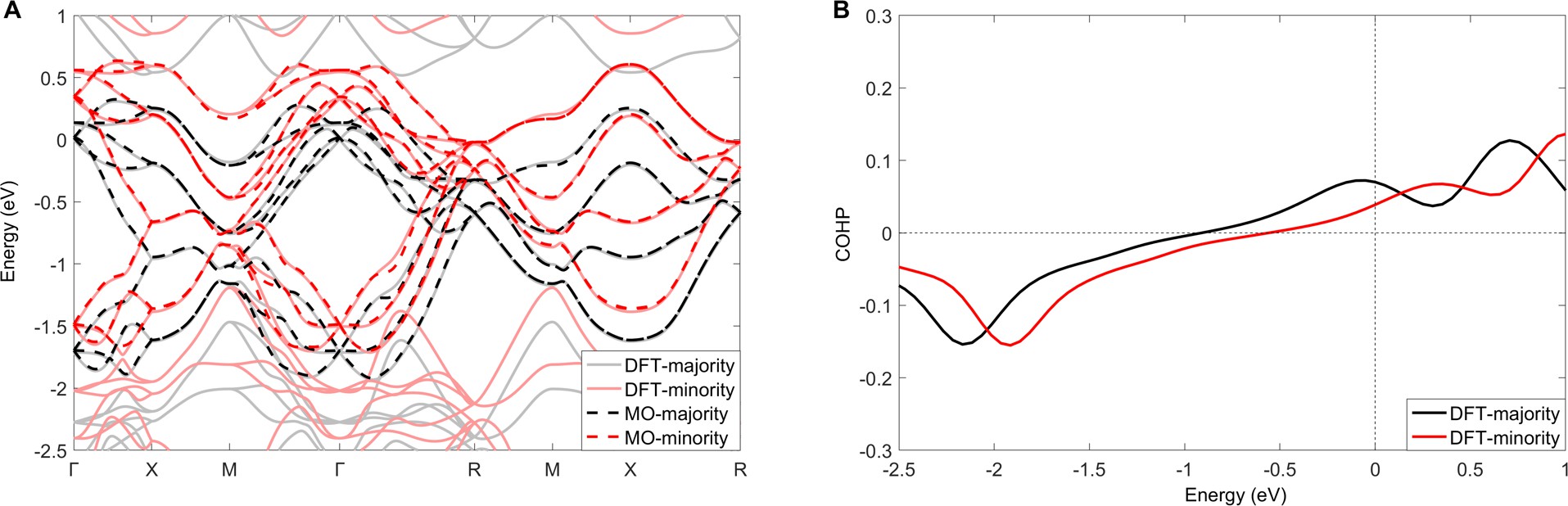}
	\caption{{\bf Electronic structure with both spin-majority and minority bands.} (\textbf{A}) DFT-calculated bands and MLWF interpolation of the 8-band EBR running through $E_\mathrm{F}$. (\textbf{B}) Spin-polarised crystal orbital Hamiltonian population (COPH) between nearest-neighbor Mn atoms.}
	\label{Figs23}
\end{figure*}
\clearpage
\begin{table}[htbp]
\centering
\begin{tabular}{|c|c|}
\hline
high-symmetry $k$ point & symmetry indicators\tabularnewline
\hline
\hline
GM & GM4 (3); GM4 (3); GM2 + GM3 (2)\tabularnewline
\hline
X & X1 (2); X1 (2); X1 (2); X1 (2)\tabularnewline
\hline
M & M3 +M4 (2); M1 + M2 (2); M3 + M4 (2); M1 + M2 (2)\tabularnewline
\hline
R & R2 + R2 (4); R1 + R3 (4)\tabularnewline
\hline
\end{tabular}
\caption{Symmetry indicators at high-symmetry $k$ points for the 8 conduction bands. Numbers in parenthesis indicate degeneracy, and the energies are sorted (from left to right) in ascending order. Spin-majority and -minority bands give the same result.}
\label{band representation}
\end{table}

\begin{table}[htbp]
\renewcommand{\arraystretch}{1.5}
\centering
\begin{tabular}{|c|c|c|c|c|c|}
\hline
\multirow{5}{*}{\includegraphics[scale=0.55]{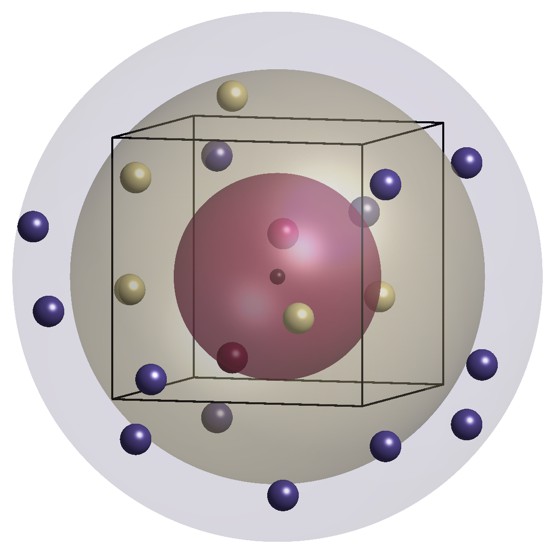}} & distance \AA & AO number & AO weight & $\eta_{\text{min}}$ within EBR  & $\eta_{\text{min}}$ outsite EBR\tabularnewline
\cline{2-6} \cline{3-6} \cline{4-6} \cline{5-6} \cline{6-6}
 & 1.78 & 10 & 0.542 & $1\times10^{-6}$ & 0.71\tabularnewline
\cline{2-6} \cline{3-6} \cline{4-6} \cline{5-6} \cline{6-6}
 & 3.6 & 61 & 0.735 & $3\times10^{-8}$ & 0.78\tabularnewline
\cline{2-6} \cline{3-6} \cline{4-6} \cline{5-6} \cline{6-6}
 & 4.6 & 147 & 0.866 & $4\times10^{-8}$ & 0.92\tabularnewline
\cline{2-6} \cline{3-6} \cline{4-6} \cline{5-6} \cline{6-6}
 & 12 & 1614 & 0.991 & $1\times10^{-7}$ & 0.97\tabularnewline
\hline
\end{tabular}
\caption{AO-projection and hopping analysis for the MOs. Rows correspond to different cut-off distance for the MO-AO projection. The origin (denoted by a small dot in the figure) is taken as the Wannier center of the MO $W_{M_{4},a}$ in the EBR $E\uparrow G$, and the first three chosen cut-off distances are displayed as transparent spheres centered at the origin, with their enclosed Mn atoms displayed as color-coded small spheres. Only Mn atoms are shown, despite some of the MO weight is contributed by Si AOs. The displayed ``AO weight'' is the sum of AO coefficient squared within the cut-off distance. A full list of $\eta$ (between $W_{M_{4},a}$ and a total of 42592 MOs, after the AO projection and truncation) is computed for each cut-off distance, and only $\eta_{\text{min}}$ connecting $W_{M_{4},a}$ to MOs within and outside the magnetic EBR are displayed. $\eta_{\text{min}}$ is used as a figure of merit for accessing the quality of the construction of the MO from AOs, see description in text.}
\label{MOhop}
\end{table}

\end{document}